\documentclass[12pt,letterpaper]{article}

\usepackage[letterpaper,margin=1in]{geometry}
\usepackage[T1]{fontenc}
\usepackage[utf8]{inputenc}
\usepackage{lmodern}
\usepackage{amsmath}
\usepackage{amssymb}
\usepackage{amsthm}
\usepackage{graphicx}
\usepackage{caption}
\usepackage{xcolor}
\usepackage{hyperref}
\usepackage{placeins}
\usepackage[capitalise,nameinlink]{cleveref}

\numberwithin{equation}{section}
\newcommand{\hideshowlabels}{}

\usepackage{framed}    % framed environment (e.g., framed, shaded)
\usepackage{xspace}    % clever spacing or not spacing
\usepackage{multicol}  % multi-column mode

\newif\ifepimomarxiv
\epimomarxivfalse

\makeatletter
\@ifundefined{ifanonymous}{%
  \newif\ifanonymous
  \anonymousfalse
}{}
\@ifundefined{iflinenumbers}{%
  \newif\iflinenumbers
  \linenumbersfalse
}{}
\makeatother

\iflinenumbers
  \usepackage{lineno}
  
  \makeatletter
  \@ifclassloaded{rsproca_new}{%
    \AddToHook{cmd/maketitle/after}[epimom/linenumbers]{\linenumbers}%
  }{%
    \@ifclassloaded{rspb_current_style}{%
      \def\epimom@makelinenumber{%
        \ifnum\value{page}=\@ne
          \makeLineNumberRight
        \else
          \makeLineNumberLeft
        \fi
      }%
      \let\makeLineNumberRunning\epimom@makelinenumber
      \AddToHook{cmd/maketitle/after}[epimom/linenumbers]{\linenumbers}%
    }{%
      \AtBeginDocument{\linenumbers}%
    }%
  }%
  \makeatother
\fi

\ifdefined\hideshowlabels
\else
  \definecolor{labelgray}{gray}{0.8}
  
  \usepackage{showlabels}
  
\fi

\newcommand{\ie}{\textit{i.e., }}
\newcommand{\eg}{\textit{e.g., }}
\newcommand{\etc}{\emph{etc.}\xspace}

\newcommand{\cf}{\textit{cf.\ }}

\newcommand{\KM}{KM\xspace}
\newcommand{\RE}{renewal equation\xspace}
\newcommand{\FoItext}{FoI\xspace}

\newcommand{\mathscale}[2]{%
  \mathchoice
    {\scalebox{#1}{$\displaystyle #2$}}%
    {\scalebox{#1}{$\textstyle #2$}}%
    {\scalebox{#1}{$\scriptstyle #2$}}%
    {\scalebox{#1}{$\scriptscriptstyle #2$}}%
}

\newcommand{\R}{{\mathcal R}}
\newcommand{\Rn}{\R_{\mathscale{\mymathscale}{0}}}
\newcommand{\Reff}{\R_{\mathscale{\mymathscale}{{\mathrm{eff}}}}}
\newcommand{\Ra}{\R_{\mathscale{\mymathscale}{\alpha}}}

\newcommand{\FoI}{F}        % force of infection (or "field of infectivity")
\newcommand{\aoi}{\alpha}   % age of infection
\newcommand{\aoidum}{\aoi'}
\newcommand{\aoiinit}{\aoi_\init}
\newcommand{\taudum}{\tau'}
\newcommand{\inc}{\iota}    % incidence
\newcommand{\cuminc}{\underline{\inc}} % cumulative incidence: lower integral
\newcommand{\gisurv}{\overline{g}}      % generation-interval survival function: upper integral

\newcommand{\Timetrans}{\mathsf{T}}
\newcommand{\Volterra}{V}

\newcommand{\Eprop}{Y_{\!\scalebox{0.5}{$E$}}}
\newcommand{\Iprop}{Y_{\!\scalebox{0.5}{$I$}}}

\newcommand{\Tlat}{T_{\rm lat}}

\newcommand{\Winv}{\mathscr{E}} % inverse of Lambert W
\newcommand{\Wp}{W_{\!\smallplus}}
\newcommand{\Wm}{W_{\!\smallminus}}
\newcommand{\Wpm}{W_{\!\smallplusminus}}

\newcommand{\Xpm}{X^{\smallplusminus}}
\newcommand{\FoIpm}{\FoI^{\smallplusminus}}
\newcommand{\rpm}{r^{\smallplusminus}}
\newcommand{\mymathscale}{0.7} % originally 0.6
\newcommand{\Xall}{X_{\mathscale{\mymathscale}{\Rn,\xinit,\yinit}}}

\newcommand{\xpmsubinit}{\xpm_{\mathscale{\mymathscale}{\xinit,\yinit}}}
\newcommand{\Xpall}{\Xall^+}
\newcommand{\Xmall}{\Xall^-}
\newcommand{\Xpmall}{\Xall^\pm}

\newcommand{\tauinit}{\tau_{\rm i}}
\newcommand{\xinit}{x_{\rm i}}
\newcommand{\yinit}{y_{\rm i}}
\newcommand{\zinit}{z_{\rm i}}

\newcommand{\epropinit}{y^{\scalebox{0.5}{$E$}}_{\mathrm{i}}}
\newcommand{\ipropinit}{y^{\scalebox{0.5}{$I$}}_{\mathrm{i}}}
\newcommand{\incinit}{\inc_{\rm i}}

\newcommand{\taupeak}{\hat{\tau}}
\newcommand{\xpeak}{\hat{x}}
\newcommand{\ypeak}{\hat{y}}

\newcommand{\smallplus}{{\mathscale{\mymathscale}{+}}}
\newcommand{\smallminus}{{\mathscale{\mymathscale}{-}}}
\newcommand{\smallplusminus}{{\mathscale{\mymathscale}{\pm}}}
\newcommand{\smallminusplus}{{\mathscale{\mymathscale}{\mp}}}
\newcommand{\xp}{x^{\smallplus}}  %{\infty}}
\newcommand{\xm}{x^{\smallminus}} %{-\infty}}
\newcommand{\xpm}{x^\smallplusminus}
\newcommand{\xmp}{x^\smallminusplus}
\newcommand{\zp}{z^{\smallplus}} %{\infty}}
\newcommand{\zptot}{\zp_{\mathscale{\mymathscale}{\mathrm{tot}}}}
\newcommand{\zm}{z^{\smallminus}} %{-\infty}}
\newcommand{\lamp}{\lambda^{\!{\smallplus}}}
\newcommand{\lamm}{\lambda^{\!{\smallminus}}}
\newcommand{\lampm}{\lambda^{\!{\smallplusminus}}}

\newcommand{\tspanel}[1]{{\bfseries\textsf{\MakeLowercase{#1}}}\xspace}
\newcommand{\pppanel}{\tspanel{C}} % phase plane panel symbol

\newcommand{\dee}{{\rm d}}
\newcommand{\dd}[2]{{\frac{\dee{#1}}{\dee{#2}}}}

\newcommand{\ddx}[1]{\dd{#1}{x}}
\newcommand{\ddt}[1]{\dd{#1}{t}}
\newcommand{\ddtau}[1]{\dd{#1}{\tau}}

\newcommand{\didi}[2]{\frac{\partial#1}{\partial#2}}

\usepackage{mathrsfs} % \mathscr
\DeclareFontFamily{U}{rsfs}{\skewchar\font127}
\DeclareFontShape{U}{rsfs}{m}{n}{%
   <-6> rsfs5
   <6-8> rsfs7
   <8-> rsfs10
}{}
\newcommand{\Lapsym}{{\mathscr{L}}}
\newcommand{\Lap}[1]{{\Lapsym}\!\left[#1\right]}

\newcommand{\Lappm}{\Lapsym_{\!\smallplusminus}}
\newcommand{\Lapm}{\Lapsym_{\!\smallminus}}
\newcommand{\Lapp}{\Lapsym_{\!\smallplus}}

\newcommand{\Oh}{{\mathcal O}}

\newcommand{\term}[1]{{\bfseries\slshape #1}}
\makeatletter

\AddToHook{cmd/appendix/after}[epimom/cleveref-appendix-types]{%
  \crefalias{section}{appendix}%
  \crefalias{subsection}{subappendix}%
  \crefalias{subsubsection}{subsubappendix}%
}

\newcommand{\beginsection}[2]{%
  \section*{#1}%
  \phantomsection%
  \def\@currentlabelname{#1}%
  \label{#2}%
  \pdfbookmark[1]{#1}{#2}%
}
\newcommand{\beginappendix}[2]{\beginsection{#1}{#2}}

\newcommand{\beginsubsection}[2]{%
  \subsection*{#1}%
  \phantomsection%
  \def\@currentlabelname{#1}%
  \label{#2}%
  \pdfbookmark[2]{#1}{#2}%
}
\newcommand{\beginsubappendix}[2]{\beginsubsection{#1}{#2}}

\newcommand{\beginsubsubsection}[2]{%
  \subsubsection*{#1}%
  \phantomsection%
  \def\@currentlabelname{#1}%
  \label{#2}%
  \pdfbookmark[2]{#1}{#2}%
}

\makeatother

\newif\ifshowappbackrefs
\showappbackrefsfalse  % uncomment for clean submission builds

\DeclareRobustCommand{\appref}[2][]{%
  \if\relax\detokenize{#1}\relax
    \cref{#2}%
  \else
    \hypertarget{appref:#1}{}\cref{#2}%
  \fi
}

\newcommand{\appbackref}[2]{%
  \hyperlink{appref:#1}{#2}%
}

\newcommand{\appbackrefs}[1]{%
  \ifshowappbackrefs
    \par\smallskip
    \noindent{\color{red}{\bfseries Referenced from: }#1.}%
    \par\smallskip
  \fi
}

\newcommand{\appfilename}[1]{%
  \ifshowappbackrefs
    \par\smallskip
    \noindent{\footnotesize\textit{Appendix source file: }\texttt{\detokenize{#1}}.}%
    \par\smallskip
  \fi
}

\makeatletter

\newcommand{\mmsectionlabel}[1]{%
  \phantomsection%
  \def\@currentlabelname{Materials and Methods}%
  \label{#1}%
  \pdfbookmark[1]{Materials and Methods}{#1}%
}

\newcommand{\mmsubsection}[2]{%
  \subsection*{#1}%
  \phantomsection%
  \def\@currentlabelname{#1}%
  \label{#2}%
  \pdfbookmark[2]{#1}{#2}%
}

\newcommand{\mmsubsubsection}[2]{%
  \subsubsection*{#1}%
  \phantomsection%
  \def\@currentlabelname{#1}%
  \label{#2}%
  \pdfbookmark[2]{#1}{#2}%
}

\renewcommand{\mmsectionlabel}[1]{%
  \phantomsection%
  \def\@currentlabelname{Appendix A}%
  \label{#1}%
  \pdfbookmark[1]{Appendix A}{#1}%
}

\renewcommand{\mmsubsection}[2]{%
  \subsection{#1}%
  \phantomsection%
  \def\@currentlabelname{#1}%
  \label{#2}%
  \pdfbookmark[2]{#1}{#2}%
}

\renewcommand{\mmsubsubsection}[2]{%
  \subsubsection{#1}%
  \phantomsection%
  \def\@currentlabelname{#1}%
  \label{#2}%
  \pdfbookmark[2]{#1}{#2}%
}

\makeatother

\let\citeref\cite

\newcommand{\nocomment}[3]{}

\usepackage[normalem]{ulem}
\newcommand{\stkout}[1]{{\color{red}\ifmmode\text{\sout{\ensuremath{#1}}}\else\sout{#1}\fi}}
\newcommand{\init}{{\mathrm{i}}}
\newcommand{\prev}{P}

\newcommand{\Rnmin}{\R_{\mathscale{\mymathscale}{0,\textrm{\upshape min}}}}

\newcommand{\Rnstar}{{\Rn^\ast}}
\newcommand{\xpeakstar}{{\xpeak_\ast}}

\usepackage{mathtools} % \coloneqq
\DeclareMathOperator*{\argmin}{arg\,min}

\newcommand{\zpmin}{\zp_{\mathscale{\mymathscale}{\vee}}}
\newcommand{\zpmax}{\zp_{\mathscale{\mymathscale}{\wedge}}}
\newcommand{\zplim}{\zp_{\mathscale{\mymathscale}{\mathrm{lim}}}}
\newcommand{\Rnstarmin}{\R^*_{\mathscale{\mymathscale}{\mathrm{0},\vee}}}
\newcommand{\Rnstarmax}{\R^*_{\mathscale{\mymathscale}{\mathrm{0},\wedge}}}

\usepackage{amssymb} % \mathbb

\epimomarxivtrue
\newcommand{\mstitle}{Epidemic ``momentum'' and a conservation law for infectious disease dynamics}

\newcommand{\msauthorone}{David J.\,D.\ Earn}
\newcommand{\msauthortwo}{Todd L.\ Parsons}

\newcommand{\msarticleauthors}{%
  \msauthorone\thanks{Department of Mathematics \& Statistics and
  M.\,G.\ DeGroote Institute for Infectious Disease Research, McMaster
  University, 1280 Main Street West, Hamilton, Ontario, L8S 4K1, Canada.
  Email: \href{mailto:earn@math.mcmaster.ca}{earn@math.mcmaster.ca}.
  ORCID: \href{https://orcid.org/0000-0002-7562-1341}{0000-0002-7562-1341}.}
  \and
  \msauthortwo\thanks{LPSM, Sorbonne Universit\'e, CNRS UMR 8001,
  Paris 75005, France. ORCID:
  \href{https://orcid.org/0000-0002-2599-8415}{0000-0002-2599-8415}.}%
}

\newcommand{\mskeywords}{epidemic momentum, epidemic modelling, final size, incidence, SIR model, SEIR model, renewal equation, conservation law, basic reproduction number, prior immunity}

\newcommand{\msfunding}{We were supported by the Fields Institute for Research in
  Mathematical Sciences; the contents of this paper are solely the
  responsibility of the authors and do not necessarily represent the
  official views of the Institute.  DJDE was supported by a Discovery
  Grant from the Natural Sciences and Engineering Research Council of
  Canada (NSERC).}
\newcommand{\msacknowledgements}{We are grateful to Ben Bolker, David Champredon, Caroline Colijn,
  Jonathan Dushoff, Maya Earn, Mark Lewis, Junling Ma, David Price,
  and Steve Walker for comments and discussions.}
\newcommand{\msaiuse}{We used ChatGPT and OpenAI Codex to assist with
  R coding, \LaTeX\ coding, and language editing.  All scientific
  arguments, mathematical derivations, analyses, interpretations, and
  conclusions were developed by the authors, who take full
  responsibility for the final manuscript.}

\newcommand{\epimomarticlemetadata}{%
  \title{\mstitle}%
  \ifanonymous
    \author{}%
    \hypersetup{%
      pdfauthor={},
      pdftitle={\mstitle},
      pdfsubject={Anonymous review manuscript},
      pdfkeywords={\mskeywords}}%
  \else
    \author{\msarticleauthors}%
    \hypersetup{%
      pdftitle={\mstitle},
      pdfsubject={},
      pdfkeywords={\mskeywords}}%
  \fi
  \date{}%
}

\newcommand{\epimomarticlekeywords}{%
  \par\medskip
  \noindent\textbf{Keywords:} \mskeywords\par
}

\renewcommand{\beginsection}[2]{\section{#1}\label{#2}}
\renewcommand{\beginsubsection}[2]{\subsection{#1}\label{#2}}
\renewcommand{\beginsubsubsection}[2]{\subsubsection{#1}\label{#2}}
\renewcommand{\beginappendix}[2]{\section{#1}\label{#2}}
\renewcommand{\beginsubappendix}[2]{\subsection{#1}\label{#2}}

\epimomarticlemetadata

\begin{document}

\maketitle

\begin{abstract}
  Infectious disease outbreaks have precipitated a profusion of mathematical models. Epidemic curves predicted by these models are typically qualitatively similar, despite distinct model assumptions, but there is no theoretical explanation for this similarity in terms of any recognised common structure. We introduce a unifying concept of \emph{epidemic momentum}---prevalence weighted by potential to infect---which is more informative than prevalence, yet analytically tractable. Epidemic momentum reveals a common underlying geometry in which outbreak trajectories always follow contours of a conserved quantity.  This previously unrecognised conservation law constrains how epidemics can unfold and provides the mathematical basis for disentangling transmissibility from prior immunity using a single epidemic time series.  Epidemic momentum also exposes the true final size of an outbreak and a universal phase-plane description that links generic renewal models to the classical SIR system.

\end{abstract}

\epimomarticlekeywords

% The shared theory main file starts with \maketitle for the formatted build.
\let\maketitle\relax

\maketitle

%%\tableofcontents

\beginsection{Introduction}{sec:intro}

Most developments in the mathematical theory of epidemics trace back
to the extremely influential contributions of Kermack and McKendrick (\KM)
\citeref{KermMcKe27,KermMcKe32,KermMcKe33}  in the early 20th century.  
The simplest model that \KM described---now known as the (basic)
susceptible-infectious-removed (SIR) model---is usually written
\begin{equation}\label{eq:SIRstandard}
  \ddt{S} \;=\; - \tfrac{\beta}{N}\, S\,I \,, \qquad
  \ddt{I} \;=\; \big(\tfrac{\beta}{N}\,S - \gamma\big) I \,,\qquad
  \ddt{R} \;=\; \gamma I \,,
\end{equation}
where $N=S+I+R$ is the (constant) total population size, and $\beta$
and $\gamma$ are the (constant) rates of transmission and recovery,
respectively.  The SIR model \labelcref{eq:SIRstandard} has had
enormous impact because it is motivated by biological mechanism, is
easy to understand, has solutions that resemble observed epidemics,
and is mathematically tractable in the sense that important features
of solutions of the model can be described with simple analytical
expressions.

A key result in \KM's first paper was an exact solution
of \cref{eq:SIRstandard} in the susceptible-removed phase
plane \cite[p.\,713]{KermMcKe27}.  \KM exploited this
phase plane solution to derive an approximation to the full temporal
dynamics of \cref{eq:SIRstandard}, which they then fit to observed
epidemic data \cite{Earn+24}.  While the exact phase plane solution is
useful pedagogically and for understanding the basic SIR model, applications
have been limited precisely because it has seemed to arise only from
the highly idealized SIR model.

\KM themselves appear to have been more interested in epidemic outcomes
than in the detailed temporal dynamics.  In their words:
\begin{quote}
%%In the course of time the epidemic may come to an end.
One of the most important problems in epidemiology is to ascertain whether
%%this termination occurs
[an epidemic can end]
only when no susceptible individuals are left, or whether the interplay of the various factors of infectivity, recovery and mortality, may result in termination, whilst many susceptible individuals are still present in the unaffected population.
\qquad--- \KM\cite[p.\,701]{KermMcKe27}
\end{quote}
\KM presented a theoretical solution \cite[p.\,708]{KermMcKe27}
in the form of what is commonly called the standard \emph{final size
relation}.  Exactly this relation has been shown to apply to models
that have much more structure than \cref{eq:SIRstandard}, using
arguments that do not involve a phase plane
solution \cite{KermMcKe27,MaEarn06,Mill2012}.  The generality of the
final size, which represents the endpoint of an epidemic, raises the
question of whether the full SIR phase plane dynamics are invariant in
some sense, and if so whether useful epidemiological inferences follow
from this insight.

Here, we show that the SIR phase plane does have a generic
counterpart, which follows after a new state variable is identified.
We first consider the next simplest model---the SEIR model, which
includes a latent or ``exposed'' stage---and show that taking the new
state variable to be \term{prevalence of infection} (either latent or
infectious) works.  Prevalence fails, however, for only slightly
richer models because equal counts of infected individuals need not
represent equal remaining infectious potential.  This failure
motivates us to define \term{epidemic momentum}, which reduces to
prevalence in the SIR and SEIR cases but reveals a universal
susceptible-momentum phase-plane geometry and a conservation law for a
broad class of infectious disease transmission models.

\beginsection{Epidemic models and phase plane solutions}{sec:models}

\beginsubsection{SIR model}{sec:SIR}

In \cref{eq:SIRstandard}, the state variables are the numbers of
individuals that are \underline{S}usceptible,
\underline{I}nfectious, or
\underline{R}emoved (recovered and immune, isolated, deceased, or
otherwise removed from the transmission process). Since the population
size $N$ is constant, the system is completely specified by only two
equations (we focus on $S$ and $I$).  Scaling the variables
to be proportions ($X=S/N$, $Y=I/N$), and measuring time in units of
the mean infectious period ($\tau=\gamma t$), the equations can be
written
\begin{subequations}\label{eq:SIR}
\begin{alignat}{2}
  \ddtau{X} &\;=\; -\Rn\,X\,Y,
  \qquad & X(\tauinit) \;&=\; \xinit,
  \label{eq:SIR;X}
  \\
  \ddtau{Y} &\;=\; (\Rn X - 1)Y,
  \qquad & Y(\tauinit) \;&=\; \yinit,
  \label{eq:SIR;Y}
\end{alignat}
\end{subequations}
where the only parameter is the \term{basic reproduction number}
($\Rn=\frac{\beta}{\gamma}$), the expected number of infections that
would be caused by a single infective individual in an otherwise fully
susceptible population.  Our notation for the initial conditions
$(\xinit,\yinit)$ follows the convention, common in probability
theory \citeref{Pars+24}, that upper case refers to functions and
lower case refers to independent variables and the values of functions
at specific points.
To be biologically admissible, the initial conditions must satisfy
\begin{equation}
  \xinit>0,
  \qquad
  \yinit>0,
  \qquad
  \xinit+\yinit\leq1 .
  \label{eq:initial.simplex.constraint}
\end{equation}
\cref{fig:phaseplane}\tspanel{A} shows $X(\tau)$
and $Y(\tau)$, obtained by numerical integration of \cref{eq:SIR} with
$\Rn=4$.

\beginsubsection{Phase plane solution}{sec:ppsol}

Eliminating time by dividing \cref{eq:SIR;X,eq:SIR;Y} yields a simple,
separable differential equation,
\begin{align}
  \ddx{Y} &\;=\; -1 + \frac{\xpeak}{x} \,,
  \qquad Y(\xinit) \;=\; \yinit, \label{eq:dYdx} \\
\intertext{where $\xpeak=\frac{1}{\Rn}$.  The exact solution is}
  Y(x) &\;=\; \yinit + (\xinit - x) - \xpeak\log{\frac{\xinit}{x}} \,.
  \label{eq:Yofx}
\end{align}
Since $Y''(x) = -{\xpeak}/{x^{2}} < 0$ for any $x>0$, the function
$Y(x)$ is strictly concave and its unique maximum is at
$(\xpeak,\ypeak)$, where $\ypeak=Y(\xpeak)$. If $\xinit<\xpeak$ then
the \emph{temporal} maximum of $Y$ occurred before the initial time
$\tauinit$.

The phase portrait $Y(x)$ \labelcref{eq:Yofx} is shown
in \cref{fig:phaseplane}\pppanel.  The function $Y(x)$ has two roots,
$\xm$ and $\xp$, which are highlighted on each trajectory in
\cref{fig:phaseplane}\pppanel.  The white points ($\xm$) correspond
to the proportion of the population that was susceptible \emph{before}
the epidemic, whereas the black dots ($\xp$) correspond to the
proportion that remain susceptible \emph{after} the epidemic.  

% Allow this large float to share a page with following text in the
% one-column arXiv presentation.
\ifepimomarxiv
  \begin{figure}[!t]
\else
  \begin{figure*}
\fi
\centerline{\bfseries Model-specific time series\hspace{3cm} Universal phase portrait}
\includegraphics[width=\textwidth]{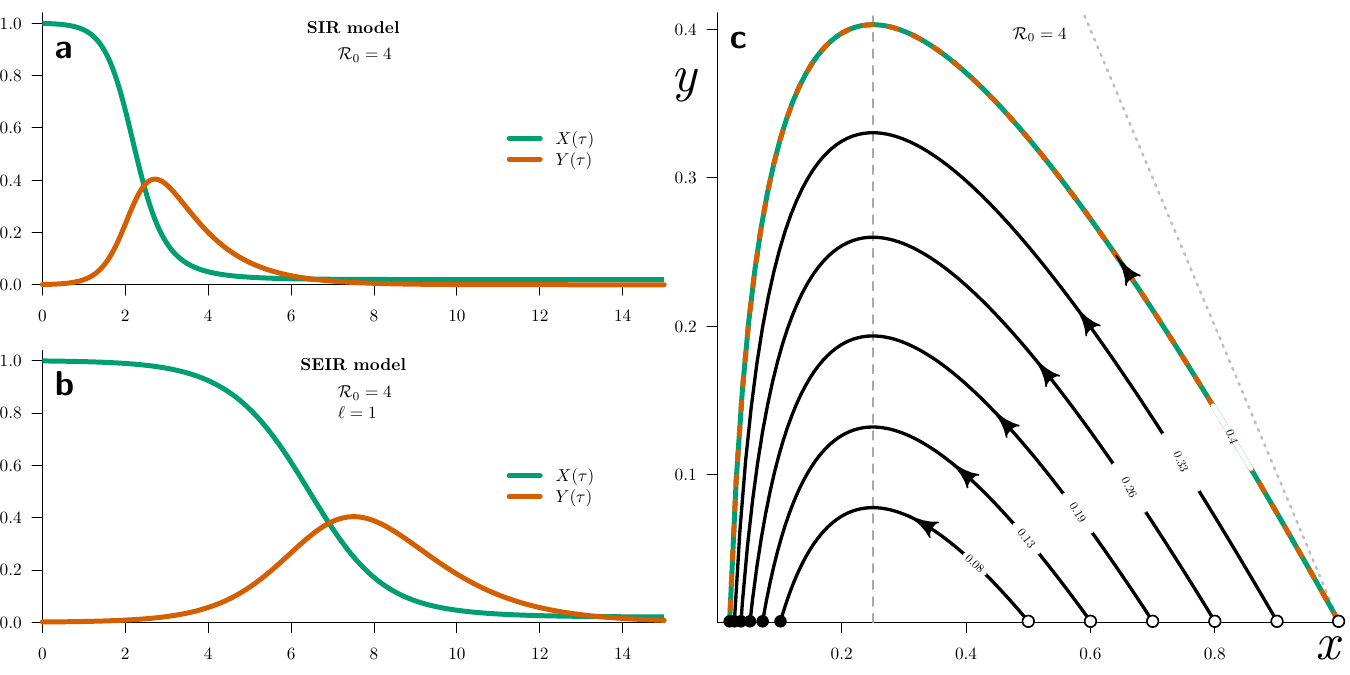}
\caption{\textbf{Universality of the susceptible-momentum ($x$-$y$)
    phase plane.}  (\tspanel{A},\tspanel{B}) Time series of the
  susceptible fraction, $X(\tau)$, and epidemic momentum, $Y(\tau)$,
  for the SIR and SEIR models [\cref{eq:SIR,eq:SEIR}] with basic
  reproduction number $\Rn=4$ and initial conditions
  $(\xinit,\yinit)=(0.999,0.00075)$.  For the SEIR model, the mean
  latent period is the same as the mean infectious period ($\ell=1$)
  and the initial exposed and infectious fractions
  [\cref{eq:SEIR;Eprop,eq:SEIR;Iprop}] are
  $\epropinit=\ipropinit=\yinit/2$.  (\pppanel) Phase portraits in the
  susceptible-momentum phase plane, which are identical for both
  models, and are expressed explicitly by \cref{eq:Yofx}.
  Trajectories are contours of constant $C(x,y)$
  [\protect\cref{eq:FI}] and are labelled with the value of the
  constant.  Both the SIR and SEIR time series on the left correspond
  to the same (coloured) phase plane trajectory on the right.  The
  dotted line is the biological boundary, $x+y=1$.  The dashed line
  indicates peak epidemic momentum, which occurs at
  $x=\xpeak=\frac{1}{\Rn}$.  The points plotted on the $x$-axis are at
  $\xm$ (white) and $\xp$ (black), the pre- and post-epidemic
  susceptible proportions for each trajectory [\cref{eq:xpm}].}
\label{fig:phaseplane}
\ifepimomarxiv
  \end{figure}
\else
  \end{figure*}
\fi

\beginsubsection{SEIR model}{sec:SEIR}

The SEIR model has two infected compartments: \term{exposed} individuals who
are not yet infectious ($\Eprop$) and \term{infectious} individuals
($\Iprop$).  Writing $\ell$ for the mean latent period in units of the
mean infectious period, the dimensionless SEIR equations are
\begin{subequations}\label{eq:SEIR}
\begin{alignat}{2}
  \ddtau{X} &\;=\; -\Rn X\Iprop,
    \qquad & X(\tauinit) &\;=\; \xinit,
    \label{eq:SEIR;X}
  \\
  \ddtau{\Eprop} &\;=\; \Rn X\Iprop - \frac{1}{\ell}\Eprop,
    \qquad & \Eprop(\tauinit) &\;=\; \epropinit,
    \label{eq:SEIR;Eprop}
  \\
  \ddtau{\Iprop} &\;=\; \frac{1}{\ell}\Eprop - \Iprop,
    \qquad & \Iprop(\tauinit) &\;=\; \ipropinit.
    \label{eq:SEIR;Iprop}
\end{alignat}
\end{subequations}
We define $Y=\Eprop+\Iprop$ so that $Y$ represents the \term{total
prevalence of infection} in both the SIR and SEIR models.
Adding \cref{eq:SEIR;Eprop,eq:SEIR;Iprop} reveals that the dynamical
equations for $X$ and $Y$ can be written identically for both models:
\begin{subequations}\label{eq:XYF}
\begin{alignat}{2}
  \ddtau{X} &\;=\; -X\FoI \,, \qquad & X(\tauinit)\;&=\;\xinit,
\label{eq:XYF;X}\\
\noalign{\vspace{5pt}}
\ddtau{Y} &\;=\; \big(X - \xpeak\big) \FoI\,, \qquad & Y(\tauinit) \;&=\; \yinit.
\label{eq:XYF;Y}
\end{alignat}
\end{subequations}
Here the \term{force of infection} 
(\FoItext) is $\FoI=\Rn Y$ for SIR and $\FoI=\Rn\Iprop$ for SEIR, and
$\xpeak=\frac{1}{\Rn}$ is the susceptible proportion when $Y$ reaches
its maximum value.  \Cref{eq:XYF} is a closed system for the
SIR model, whereas a third equation (\cref{eq:SEIR;Eprop}
or \labelcref{eq:SEIR;Iprop}) is needed to complete the specification
of the SEIR model.

The key insight from \cref{eq:XYF} is that $\ddx{Y}$ does not
depend on the \FoItext $\FoI$, so we obtain the same
susceptible-prevalence phase plane equation \labelcref{eq:dYdx} and
solution \labelcref{eq:Yofx} for both the SIR and SEIR models (hence
the identical phase portrait in spite of different time series for SIR
and SEIR in \cref{fig:phaseplane}).  To our knowledge, this phase
portrait invariance has not been noticed previously.

\beginsubsection{SI$_1$I$_2$R model}{sec:SI1I2R}

Prevalence is not, however, the right state variable to complement the
susceptible fraction in general.  The problem is immediately evident
if we consider the simplest extension of the SEIR model, replacing $E$
and $I$ with infectious stages $I_1$ and $I_2$.  Using proportions, we
label the infected variables $Y_1$ and $Y_2$ and, for simplicity, we
assume that the two stages have the same transmissibility and mean
duration, which yields an Erlang SI$^2$R model \cite{KrylEarn13},
\begin{subequations}
\label{eq:SI1I2R}
\begin{alignat}{2}
  \ddtau{X} &\;=\; -\Rn X (Y_1 + Y_2),
  \qquad && X(\tauinit) \;=\; \xinit,
  \label{eq:SI1I2R;X}
  \\
  \ddtau{Y_1} &\;=\; \Rn X (Y_1 + Y_2) - Y_1,
  \qquad && Y_1(\tauinit) \;=\; y_{1,\init},
  \label{eq:SI1I2R;Y1}
  \\
  \ddtau{Y_2} &\;=\; Y_1 - Y_2,
  \qquad && Y_2(\tauinit) \;=\; y_{2,\init}.
  \label{eq:SI1I2R;Y2}
\end{alignat}
\end{subequations}
The total prevalence is
\begin{equation}
  \prev \;=\; Y_1 + Y_2,
  \label{eq:SI1I2R;prev}
\end{equation}
and the sum of \cref{eq:SI1I2R;Y1,eq:SI1I2R;Y2} is
\begin{equation}
  \ddtau{\prev}
  \;=\;
  \Rn X \prev - Y_2.
  \label{eq:SI1I2R;prevdot}
\end{equation}
Combining \cref{eq:SI1I2R;prevdot,eq:SI1I2R;X} yields
\begin{equation}
  \ddx{\prev}
  \;=\;
  -1 + \frac{1}{\Rn x}\Big(\frac{Y_2}{\prev}\Big).
  \label{eq:SI1I2R;prevalence.phase}
\end{equation}
Thus the prevalence-based phase-plane structure breaks even for this
simple $SI_1I_2R$ model.  The susceptible-prevalence phase-plane
equation \labelcref{eq:SI1I2R;prevalence.phase} is \emph{not} closed;
it depends not only on $\prev$ and $x$ but on the proportion of infectious
individuals who are in each infectious stage (through the ratio
$Y_2/P$).

The dependence of \cref{eq:SI1I2R;prevalence.phase} on how prevalence
is distributed among infectious stages is an important clue that leads
us to define the infectious state variable that reveals a universal
phase plane geometry for epidemics.  Rather than simply counting
currently infectious individuals (prevalence), we need
to \emph{weight} individuals according to their potential to infect in
the future.  To make this precise, we first consider a generic
formulation of epidemic models.

\beginsubsection{Renewal equation}{sec:re}

\KM derived the SIR model as a special case of a much more general
integro-differential equation, equivalent to what is now commonly
called the \term{\RE} \citeref{KermMcKe27,Bred+12,Cham+18}.
The state variables for the \RE
are $X$, the susceptible fraction, and $\FoI$, 
the \term{force of infection} 
(\FoItext); the \FoItext is the
instantaneous hazard of
infection per susceptible individual.
The \RE is more general than ordinary differential equation
epidemic models because it allows
infectiousness to vary continuously as a function of an
individual's \term{age of infection} $\aoi$, the amount of time that
has elapsed since they were initially infected
(which may include latent and/or carrier periods when they were not infectious).
Unlike the simple compartmental models discussed above, prevalence
(the proportion of the population that is currently infected, whether
infectious or not) is not an explicit
variable\footnote{The renewal equation was derived by \KM \cite{KermMcKe27} assuming the dependence of the recovery rate on age-of-infection is known; from that relationship, one can obtain an explicit expression for the prevalence.  In practice, however, only (a proxy for) the generation interval––not the recovery rate––is observed, and separating the generation interval into recovery rate and age-of-infection-specific transmission rate is not possible without additional data.}.
%%\citenote{note:RE.prevalence}.

In the \RE framework, different models (\eg involving multiple
infectious stages, hospitalization, treatment, relapse, \etc) are
specified through the probability density, $g(\aoi)$, of
the \term{intrinsic generation interval} (the time difference between
the moment when a focal individual was infected and the earlier time
when the infector was infected \citeref{Sven07,ChamDush15}).
In dimensionless time, the governing equations are
\begin{subequations}\label{eq:re}
\begin{align}
  \ddtau{X} &\;=\; -X(\tau)\FoI(\tau), \label{eq:re;X}\\
  \FoI(\tau) &\;=\;
    \Rn\int_0^\infty
    X(\tau-\aoi)\FoI(\tau-\aoi)g(\aoi)\,\dee\aoi,
    \label{eq:re;FoI}
\end{align}
\end{subequations}
which can alternatively be expressed using the \term{incidence} of infection,
\begin{equation}\label{eq:re;inc}
  \inc \;=\; X\,\FoI,
\end{equation}
rather than the \FoItext directly.  The generation interval
distributions for the SIR and SEIR models are listed
in \cref{tab:GI.properties} and are shown to yield these models
in \appref[re-to-seir-gi]{app:re.to.seir}.  For the SIR, $g(\aoi)$ is exponential and for
the SEIR $g(\aoi)$ is a linear combination of exponentials.  Explicit
expressions for $g(\aoi)$ for many other compartmental models have
also been derived \cite{Cham+18}.

To specify an initial-value problem for \cref{eq:re}, it is not
sufficient to prescribe only the susceptible fraction at the initial
time ($\xinit$).  Because the renewal term (the integral
in \cref{eq:re;FoI}) depends on infections that occurred before the
initial time ($\tauinit$), we must also prescribe the
\term{initial history}, defined for all $\aoiinit\geq0$ by
\begin{align}
  \inc_\init(\aoiinit)
  \;&=\;
  \inc(\tauinit-\aoiinit) \notag\\
  \;&=\;
  X(\tauinit-\aoiinit)\FoI(\tauinit-\aoiinit),
  \label{eq:re.initial.history}
\end{align}
where $\aoiinit$ denotes infection age at time $\tauinit$.
Equivalently, we must specify the distribution of infection ages
$\aoiinit$ at time $\tauinit$ among individuals who can generate
future infections.  At a time $\tau\geq\tauinit$, infections acquired
between $\tauinit$ and $\tau$ have ages $0\leq\aoi\leq\tau-\tauinit$,
whereas an infection represented in the initial history with age
$\aoiinit$ at time $\tauinit$ has age $\tau-\tauinit+\aoiinit$ at time
$\tau$.  Splitting the renewal integral at $\aoi=\tau-\tauinit$ then
gives
\begin{equation}
  \FoI(\tau)
  \;=\;
  \Rn
  \Bigg[
    \int_0^\infty \! \inc_\init(\aoiinit)
      g(\tau-\tauinit+\aoiinit)\,\dee \aoiinit
      \,+\,
    \int_0^{\tau-\tauinit} \!
      X(\tau-\aoi)\FoI(\tau-\aoi)g(\aoi)\,\dee\aoi
  \Bigg].
  \label{eq:re.initial.FoI}
\end{equation}
Thus the initial data consist of $X(\tauinit)=\xinit$, together with
the non-negative initial history $\inc_\init(\aoiinit)$.  For
special choices of $g$, such as the exponential distribution that
yields the SIR model, this history can be summarized by a finite
number of compartment variables (\cf \cite{Cham+18}); for a general
renewal equation it cannot.

\beginsection{Epidemic momentum}{sec:epimom.etc}

\beginsubsection{Definition of epidemic momentum}{sec:epimom}

In the framework of the \RE we can define the quantity that,
together with susceptible fraction $X$, yields the phase plane
equation \labelcref{eq:dYdx} generically.

\hypertarget{tmp:gi}{}
In the \RE, dependence on infection age $\aoi$ is
represented by the generation interval distribution, 
\ie the probability density, $g(\aoi)$, of the
infection age at which a potential transmission might occur, ignoring
depletion of susceptibles.
We denote the survival function of the intrinsic generation interval
distribution by
\begin{equation}\label{eq:GI.survival}
  \gisurv(\aoi)
  \;\equiv\;
  \int_{\aoi}^{\infty}g(\aoidum)\,\dee\aoidum \, .
\end{equation}
At infection age $\aoi$, $\gisurv(\aoi)$ is the fraction of an
infected individual's potential reproductive output that remains.
Consequently, the individual's expected reproductive output after that
infection age, ignoring depletion of susceptibles, is
\begin{equation}\label{eq:Ra}
  \Ra \;=\; \Rn\,\gisurv(\aoi) \, .
\end{equation}
We refer to $\Ra$ as\footnote{The reduced reproduction number $\Ra$ is
closely related to Fisher's reproductive value
(see \eg \citeref[\S8.1]{KeyfCasw2005}). Unlike the reproductive
value, the reduced reproduction number is not discounted for an
exponentially growing population, but is normalized to an individual's
potential total output.}
%%\citenote{note:FisherRV}
the \term{reduced
reproduction number} at infection age $\aoi$.  The notation $\Ra$ is
chosen so that $\aoi=0$ corresponds to the basic reproduction number
$\Rn$.

At time $\tau$, individuals infected at time $\tau-\aoi$ have
infection age $\aoi$.  Therefore, the total potential for future
transmission by all currently infected individuals is
\begin{equation}\label{eq:re.Y}
  Y(\tau) \;=\; \int_{0}^{\infty} \inc(\tau-\aoi)
  \gisurv(\aoi)\, \dee \aoi \,.
\end{equation}
For the SIR and SEIR models,
we show in \appref[re-to-seir-momentum]{app:re.to.seir} that $Y$ as defined
by \cref{eq:re.Y} is exactly the total prevalence of infection.
The reason that $Y$ reduces to prevalence in these special cases is
that they unrealistically assume constant infectiousness throughout an
exponentially distributed infectious period, \ie the SIR and SEIR
models assume that the rate at which infectious individuals transmit
to others does not depend on how long they have been infectious.

Since $Y$ is equivalent to prevalence only in special cases, we need a
different term for it.  Inspired by a physics analogy that we discuss
in \cref{sec:physics.analogies}, we refer to $Y$ as the
\term{epidemic momentum}.

Setting $s=\tau-\aoi$, we can alternatively write
\begin{equation}\label{eq:re.Y.history}
  Y(\tau)
  \;=\;
  \int_{-\infty}^{\tau}
    \inc(s)\gisurv(\tau-s)\,\dee s.
\end{equation}
%%Recalling \cref{eq:GI.survival} and applying Leibniz's rule for differentiation under the integral sign to \cref{eq:re.Y.history} yields
Because $g$ is a probability density,
\cref{eq:GI.survival} implies that $\gisurv(0)=1$ and
$\dd{\gisurv}{\aoi}=-g(\aoi)$.  Differentiating
\cref{eq:re.Y.history}, we find
\begin{align}
  \ddtau{Y}
  &\;=\;
  \inc(\tau)\,\gisurv(0)
  +
  \int_{-\infty}^{\tau}
    \inc(s)\,
    \didi{}{\tau}\Big(\gisurv(\tau-s)\Big)\,\dee s
    & \text{(Leibniz's rule)}
  \notag\\
  &\;=\;
  \inc(\tau)
  -
  \int_{-\infty}^{\tau}
    \inc(s)g(\tau-s)\,\dee s
    & \text{\cref{eq:GI.survival}}
  \notag\\
  &\;=\;
  \inc(\tau)
  -
  \int_0^\infty
    \inc(\tau-\aoi)g(\aoi)\,\dee\aoi
    & (\aoi = \tau-s)
  \notag\\
  &\;=\;
  X(\tau)\FoI(\tau) - \frac{1}{\Rn}\FoI(\tau)
  & \text{\cref{eq:re;FoI,eq:re;inc}}
  \notag\\
  &\;=\;
  \bigl(X(\tau)-\xpeak\bigr)\FoI(\tau),
  & (\xpeak=\frac{1}{\Rn})
  \label{eq:re.Ydot}
\end{align}
which matches \cref{eq:XYF;Y}.
In \appref[yinit-from-re]{app:y.equiv}, we show that
the initial epidemic momentum
in \cref{eq:XYF;Y} can be expressed in terms of the
initial history \labelcref{eq:re.initial.history},
\begin{equation}\label{eq:re;yinit}
  \yinit \;=\; \int_{0}^{\infty} \inc(\tauinit-\aoi)
  \gisurv(\aoi)\, \dee \aoi.
\end{equation}
Thus the integral definition of $Y$ \labelcref{eq:re.Y}, 
together with the renewal equation \labelcref{eq:re},
implies the ODE for $Y$ \labelcref{eq:XYF;Y}.  
Conversely, in \appref{app:y.equiv},
we show that \cref{eq:XYF;Y,eq:re} together imply \cref{eq:re.Y}.
\Cref{eq:re;yinit} expresses $\yinit$ in terms of the entire initial
history, but we show in \appref[finite-int-reps-main]{app:yinit.from.incinit} that it is sufficient
for observations to begin during the exponential growth phase.
For $\xinit>\xpeak$, the subsequent trajectory reaches
$x=\xpeak=\frac{1}{\Rn}$ if and only if $\yinit>0$, and $Y$ attains
its maximum for $\tau\ge\tauinit$ at this point
[\appref[momentum-peak-existence]{app:momentum.peak}].

Given an assumed or estimated generation interval distribution
and an observed incidence curve, \cref{eq:re.Y} allows us to compute the
epidemic momentum through time.
Moreover, we show in \appref[proof-y-recover]{app:y.equiv} that given
the epidemic momentum, we can always recover the \FoItext and
incidence. Thus, momentum is an analytically tractable, explicitly
computable quantity that is interchangeable with commonly used
descriptors of epidemic dynamics.

In the remainder of this paper, we exploit the epidemic momentum to
uncover generic properties of epidemic models that can be expressed
with the \RE \labelcref{eq:re}.
We obtain new insights concerning
the basic reproduction number,
the level of population immunity before an epidemic,
the proportion of the population infected during an outbreak (the
``final size''),
and the relationship between solutions of generic epidemic models and
the simple SIR model.

\beginsubsection
    %%[Universality\dots]
    {Universality of the SIR phase portrait}{sec:phaseplane}

Combining \cref{eq:re;X,eq:re.Ydot}, we obtain exactly the
system \labelcref{eq:XYF} that we derived in the context of the
SEIR model.  Thus, taking $Y$ to be the epidemic momentum, rather than
prevalence, the classical SIR phase plane solution shown
in \cref{fig:phaseplane}\pppanel is revealed to be a universal
susceptible-momentum phase plane description of the trajectories of
any epidemic model that can be cast as a \RE \labelcref{eq:re}.

If we obtain $X(\tau)$ by numerical integration of the \RE, or any
other means, we can immediately compute $Y(\tau)$ from the phase plane
solution \labelcref{eq:Yofx}.  Conversely, if we obtain $Y(\tau)$ by
some means, such as computing it from observed incidence
[\cref{eq:re.Y}], and we assume the system can be represented by
the \RE \labelcref{eq:re}, then we can immediately reconstruct the
susceptibles $X(\tau)$ (an explicit expression for $X(y)$ is given
below in \cref{eq:Xofy}).

\beginsubsection
    %%[Equivalence\dots]
    {Equivalence of generic and SIR epidemics via time transformation}{sec:timechange}

The susceptible fraction $X(\tau)$ and the epidemic momentum
$Y(\tau)$ can be mapped via a time reparameterization onto the
trajectories of the standard SIR model \labelcref{eq:SIR}.  If we set
\begin{equation}\label{eq:timechange}
\Timetrans(\tau) \;=\; \int_{0}^{\tau} \frac{\xpeak\,\FoI(s)}{Y(s)}\, \dee s \,,
\end{equation}
then the pair $\big(X(\Timetrans^{-1}(\tau)),\;Y(\Timetrans^{-1}(\tau))\big)$
satisfies the SIR equations \labelcref{eq:SIR}, as we show
in \appref[timechange-sir-equivalence]{app:timechange}.

The most important consequence of \cref{eq:timechange} is that the
only effect of model structure more complicated than that of the
standard SIR model is to change the speed with which the geometrically
invariant solutions \labelcref{eq:Yofx} in the susceptible-momentum
phase plane (\cref{fig:phaseplane}) are traversed.

\beginsubsection
    %%[A first integral\dots]
    {A first integral for generic epidemics}{sec:FI}

Writing $y=Y(x)$ and rearranging the phase-plane
equation \labelcref{eq:Yofx} so that the initial state and general state
are separated, we have
\begin{equation}\label{eq:FI.derivation}
y + (x-\xpeak) - \xpeak\ln{\frac{x}{\xpeak}}
\;=\; \yinit + (\xinit-\xpeak) - \xpeak\ln{\frac{\xinit}{\xpeak}}.
\end{equation}
Thus, this expression has the same value at all points $(x,y)$ along a given
solution in the susceptible-momentum phase plane.  We therefore
have a \term{first integral} \citeref{Stro18} for a \emph{generic epidemic},
\begin{align}\label{eq:FI}
    C(x,y)
    \;=\; y + \xpeak\,\Volterra\big(\frac{x}{\xpeak}\big)
    \,,
\end{align}
where
\begin{equation}\label{eq:Volterra}
\Volterra(u) \;=\; u-1-\ln{u}
\end{equation}
is the ``Volterra function'' \citeref{EarnMcCl25} that arises in
global stability analyses of population models.  Since the value
of $C(x,y)$ is conserved along any trajectory, we can evaluate it
at the point of peak momentum to find
\begin{equation}\label{eq:FI.value}
C\big(X(\tau),Y(\tau)\big) \;=\; C(\xpeak,\ypeak) \;=\; \ypeak.
\end{equation}
Thus the value of the conserved quantity $C(x,y)$ is the peak
epidemic momentum $\ypeak$.

\goodbreak

\beginsubsection{Prior immunity and final size from the universal phase portrait}{sec:pp.pi.fs}

\beginsubsubsection{Explicit expressions}{sec:explicit.xpm}

Explicit expressions for $\xpm$ (the two roots of $Y(x)$) can be
obtained by inverting \cref{eq:Yofx} using Lambert's $W$ function
(see \appref[lambert-w]{app:LambertW})
and then evaluating the result at $y=0$.  The inverse of $Y(x)$ has
two branches \citeref{ParsEarn24}.  For a state with $y=Y(\tau)$, we
display their dependence on $\Rn$ and the initial conditions
$(\xinit,\yinit)$ by writing
\begin{subequations}\label{eq:Xofy}
\begin{align}
  \Xall(y) &\;=\; 
  \begin{cases}
    \Xmall(y) \,, & \text{if } \tau \le \taupeak,\\
    \noalign{\vspace{5pt}}
    \Xpall(y) \,, & \text{if } \tau > \taupeak,
  \end{cases}
\intertext{and find}
  \Xpmall(y) &\;=\;
  -\frac{1}{\Rn}\,\Wpm(-{\Rn\xinit}\,e^{-\Rn\xinit}\,e^{\Rn(y-\yinit)}) \,.
  \label{eq:Xpmofy}
\end{align}
\end{subequations}
Because $\Wm(z)\le-1$ and $\Wp(z)\ge-1$ on the relevant real
interval, $\Wm$ gives the pre-peak branch
$\Xmall(y)\ge1/\Rn$, whereas $\Wp$ gives the post-peak branch
$\Xpall(y)\le1/\Rn$.
The pre- and post-epidemic susceptible proportions are then
\begin{equation}\label{eq:xpm}
\xpmsubinit(\Rn) \;=\; \Xpmall(0) \,.
\hypertarget{xpm.ineq}{}
\end{equation}
We normally suppress the parameter dependence and write simply $X(y)$, $\Xpm(y)$ and $\xpm$; where explicit dependence of $\xpm$ on $\Rn$ is needed, we write $\xpm(\Rn)$.
For the remainder of this paper we assume the usual situation in which
the initial time $\tauinit$ is before the peak
($\xinit>\xpeak=1/\Rn$).

The \term{prior population immunity}, \ie the level of population immunity in the
population before the epidemic, is the proportion of the population
that was immune in the limit $\tau\to-\infty$,
\begin{equation}\label{eq:zm}
\zm \;=\; 1 - \xm,
\end{equation}
and the \term{final size} of the epidemic, \ie the proportion of the
population infected \emph{during} the outbreak, is
\begin{equation}\label{eq:zp}
\zp \;=\; \xm - \xp.
\end{equation}
Classical final-size relations express the asymptotic
post-epidemic susceptible fraction $\xp$ in terms of an epidemic state
specified at a finite initial
time \citeref{KermMcKe27,MaEarn06,Mill2012}. \Cref{eq:Xofy,eq:xpm}
determine both asymptotic endpoints ($\xm$ and $\xp$) from $\Rn$ and
any state on the trajectory.  Thus prior population immunity
need not be specified \emph{a priori}, and \cref{eq:zp} gives the
proportion infected during the focal epidemic.

\beginsubsubsection{Relationship to na\"ive final size}{sec:naive.final.size}

It follows from
\cref{eq:Xofy,eq:xpm} that
\begin{equation}\label{eq:R.vs.Reff.identity}
  \xm \xp_{\mathscale{\mymathscale}{1,0}}(\Rn\xm)
  \;=\;
  -\frac{1}{\Rn} \Wp(-\Rn\xm e^{-\Rn\xm})
  \;=\;
  \xp_{\mathscale{\mymathscale}{\xm\!,0}}(\Rn) ,  
\end{equation}
and hence
\begin{equation}\label{eq:zp.vs.naive}
  \zp
  \;=\;
  \xm - \xp
  \;=\;
  \xm \bigl(1-\xp(\Reff)\bigr),
\end{equation}
where $\Reff=\Rn\xm$ is the effective reproduction number in the limit
$\tau\to-\infty$. \Cref{eq:zp.vs.naive} expresses the genuine final
size $\zp$ as the genuine pre-epidemic susceptible proportion,
$\xm=\xm(\Rn)$, times the na\"ive final size, $1-\xp(\Reff)$, \ie the
final size calculated using $\Reff$ rather than the genuine $\Rn$.
Thus the na\"ive approach yields the final size relative to the
pre-epidemic susceptible population.

\beginsubsubsection{Constraints on initial conditions and final size}{sec:constraints.final.size}

The constraints on initial conditions given
in \cref{eq:initial.simplex.constraint} are necessary but not
sufficient.  For a given $\Rn$ (or, equivalently, a given $\xpeak$),
any biologically admissible initial state must yield $\xm\le1$, \ie
the phase plane trajectory that contains $(\xinit,\yinit)$
must cross
the $x$-axis ($y=0$) for some ``earlier'' $x\in(\xinit,1]$.  This means that
in \cref{eq:Yofx} we must have $Y(1)\le0$, \ie
\begin{equation}\label{eq:yinit.constraint}
  \xinit
  + \yinit
  + \xpeak\log\!\Big(\frac{1}{\xinit}\Big)
  \;\leq\;
  1 ,
\end{equation}
which is a more stringent condition than the third inequality
in \cref{eq:initial.simplex.constraint} because the logarithmic term
is positive for $0<\xinit<1$ (so, in particular,
$\xinit+\yinit<1$).  \Cref{eq:initial.simplex.constraint,eq:Xofy,eq:xpm,eq:yinit.constraint}
imply that $\xinit$ must lie in the interval $[\xp,\xm]$ and (since we have assumed
$\xinit\ge\xpeak$) that
\begin{equation}\label{eq:xp.lt.xm}
0 \;<\; \xp \;<\; \xpeak \;\le\; \xinit \;<\; \xm \;\le\; 1.
\end{equation}
The first inequality ($0 < \xp$) answers the question posed in the \KM
quote in the introduction [\cref{sec:intro}], for the generic setting of
the \RE \labelcref{eq:re}, via the universal phase portrait.

\beginsubsubsection{Constraint on $\Rn$ from given initial conditions}{sec:constraint.Rn}

If we solve inequality \labelcref{eq:yinit.constraint} for $\xpeak$, we find that
for a \emph{given} initial state $(\xinit,\yinit)$, the basic reproduction number $\Rn=1/\xpeak$ 
must satisfy
\begin{equation}\label{eq:Rnmin}
  \Rn
  \;\geq\;
  \Rnmin
  \;=\;
  \frac{\log(1/\xinit)}{1-\xinit-\yinit} .
\end{equation}
For a given initial state, \cref{eq:Rnmin} therefore gives a lower
bound on $\Rn$ that is required for biological admissibility.

\beginsubsubsection{Non-monotonicity of final size}{sec:non-monotonicity.final.size}

\begin{figure*}
\includegraphics[width=\textwidth]{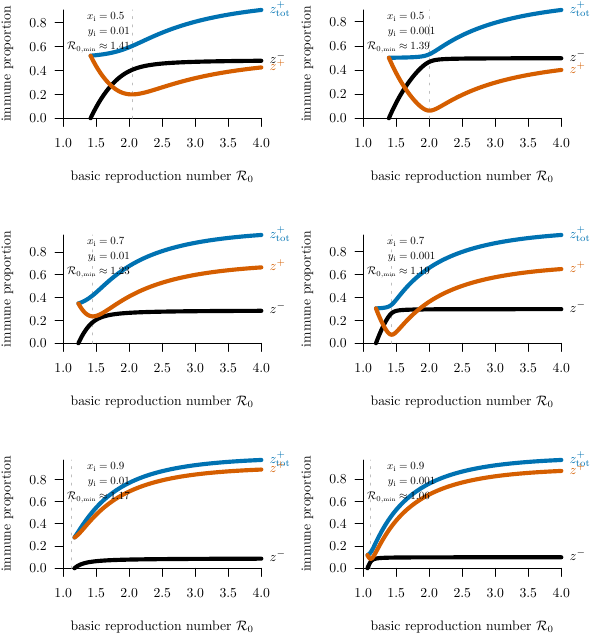}
\caption{\textbf{Prior immunity $(\zm)$ {\upshape[\cref{eq:zm}]}, total post-epidemic immune fraction $(\zptot=1-\xp)$ and final size $(\zp)$ {\upshape[\cref{eq:zp}]} vs $\Rn$ for six given initial states.}
  The vertical dotted grey line is at $\Rn={1}/({\xinit-\yinit})$, which closely approximates the
  reproduction number at which the final size is minimized {\upshape[\cref{eq:zpmin.Rnstar.leading.approx}]}.
  Non-monotonicity of the final size is possible because the initial conditions are fixed as $\Rn$
  is increased (see \cref{sec:non-monotonicity.final.size}).
  For $(\xinit,\yinit)=(0.9,0.01)$ (bottom left panel), the exact
  critical-point condition for $\zp(\Rn)$ has two solutions,
  $\Rnstar\approx1.124$ and $\Rnstar\approx9.38$.  The leading
  approximation \labelcref{eq:zpmin.Rnstar.leading.approx}, indicated
  in the figure, agrees with the exact lower critical point to four
  significant digits.  However, this point lies below
  $\Rnmin\approx1.171$ {\upshape[\cref{eq:Rnmin}]}, so it would imply a
  pre-epidemic susceptible fraction greater than one.  The second
  critical point is a large-$\Rn$ local maximum, beyond which
  $\zp(\Rn)$ decreases to $\xinit+\yinit=0.91$.  See
  \cref{tab:zp.critical.values}.}
\label{fig:zpm}
\end{figure*}

In \appref[zp-min-main]{app:zp.min} we show that, given any initial
state $(\xinit,\yinit)$, the pre-epidemic immune fraction ($\zm$) and
the total post-epidemic immune fraction ($\zptot$) strictly increase
as a function of $\Rn$, whereas the genuine final size ($\zp$) need
not be monotone as a function of $\Rn$.  We also find an implicit
expression for the exact minimum point of $\zp(\Rn)$
[\cref{eq:zpmin.minimum.condition}], and derive an accurate
approximation,
\begin{equation}\label{eq:zpmin.Rnstar.leading.approx}
  \Rnstarmin
  \;\coloneqq\;
  \argmin\!\big( \zp(\Rn) \big)
  \;\approx\;
  \frac{1}{\xinit-\yinit}.
\end{equation}
For a sample of given initial states, \cref{fig:zpm} shows, as a
function of $\Rn$ (for $\Rn>\Rnmin$), the pre-epidemic immune fraction
($\zm=1-\xm$) the total post-epidemic immune fraction ($\zptot=1-\xp$,
traditionally taken to be the final size), and the genuine final size
($\zp=\xm-\xp$, the fraction of the population that becomes
immune \emph{during} the epidemic).

A local minimum in $\zp(\Rn)$ is surprising because it means that there
is an interval of $\Rn$ values over which increasing $\Rn$
\emph{reduces} the final size of the outbreak.  This counter-intuitive
behaviour is possible because we are conditioning on a fixed state
$(\xinit,\yinit)$ at the chosen initial time $\tauinit$.  The final
size can be decomposed as
\begin{equation}
  \zp
  \;=\;
  \xm-\xp
  \;=\;
  (\xm-\xinit)+(\xinit-\xp),
  \label{eq:zp.before.after.tauinit}
\end{equation}
where $\xm-\xinit$ is the fraction infected before $\tauinit$ and
$\xinit-\xp$ is the fraction infected after $\tauinit$.  Increasing
$\Rn$ increases the transmission potential from the fixed state
$(\xinit,\yinit)$, so it increases the number infected after
$\tauinit$, thereby reducing $\xp$.  But the same larger value of
$\Rn$ also implies that fewer susceptible individuals were needed
before $\tauinit$ to reach the fixed initial state, reducing the
corresponding pre-epidemic susceptible fraction $\xm$.  On intervals
where $\zp$ decreases as $\Rn$ increases, the decrease in $\xm-\xinit$
is larger than the increase in $\xinit-\xp$, so---\emph{given the
fixed state at time $\tauinit$}---the genuine final size $\zp=\xm-\xp$
shrinks as a proposed $\Rn$ is increased, even though transmission
would be stronger.

There is also a local maximum in $\zp(\Rn)$, but it occurs for
relatively large $\Rn$ and is typically very close to $\xinit+\yinit$,
which is the limit of $\zp$ as $\Rn\to\infty$
(see \cref{tab:zp.critical.values}).

\begin{table}[h]
\centering
\caption{Critical points and limiting values of $\zp(\Rn)$ for the
initial conditions used in \cref{fig:zpm}.  The lower bound
$\Rnmin$ is the value at which $\xm=1$.  The quantities $\Rnstarmin$ and
$\Rnstarmax$ denote the local minimum and local maximum points of
$\zp(\Rn)$, respectively.  The limiting value is
$\zplim=\xinit+\yinit=\lim_{\Rn\to\infty}\zp(\Rn)$.  The dagger marks
the case in which the local minimum lies below $\Rnmin$, so it would
imply $\xm>1$.}
\label{tab:zp.critical.values}
\begin{tabular}{cccccccc}
\hline
$\xinit$
&
$\yinit$
&
$\Rnmin$
&
$\Rnstarmin$
&
$\zpmin$
&
$\Rnstarmax$
&
$\zpmax$
&
$\zplim$
\\
\hline
$0.5$ & $0.01\phantom{0}$
&
$1.4146$ & $2.0415$ & $0.1992$ & $15.1353$ & $0.5113$ & $0.5100$
\\
$0.5$ & $0.001$
&
$1.3891$ & $2.0040$ & $0.0632$ & $21.5038$ & $0.5011$ & $0.5010$
\\
$0.7$ & $0.01\phantom{0}$
&
$1.2299$ & $1.4495$ & $0.2360$ & $11.5380$ & $0.7112$ & $0.7100$
\\
$0.7$ & $0.001$
&
$1.1929$ & $1.4306$ & $0.0748$ & $15.9691$ & $0.7011$ & $0.7010$
\\
$0.9$ & $0.01\phantom{0}$
&
$1.1707$ & $\phantom{{}^\dagger}1.1237^\dagger$ & $0.2677$ & $\phantom{1}9.3818$ & $0.9112$ & $0.9100$
\\
$0.9$ & $0.001$
&
$1.0642$ & $1.1123$ & $0.0848$ & $12.7702$ & $0.9011$ & $0.9010$
\\
\hline
\end{tabular}
\end{table}

\FloatBarrier

%%\beginsubsection{New insights from the rise and fall of outbreaks}{sec:rise.fall}
\beginsubsection{Rising and falling tail exponents, $\lampm$}{sec:rise.fall}

In \appref[tail-exponents-main]{app:lampm.re}, we show that epidemic
momentum $Y$, \FoItext $\FoI$, and incidence $\inc$ always have the
same asymptotic exponential growth rate, $\lamm>0$, as
$\tau\to-\infty$.  If $g(\aoi)$ vanishes beyond some finite infection
age (as must always occur in real biological systems), $Y$, $\FoI$ and
$\inc$ also decay exponentially at an identical rate $\lamp<0$ as
$\tau\to+\infty$ (\appref{app:lampm.re.exist} gives a more general
sufficient condition for existence of $\lamp$ and an unbounded-support
counterexample).  We assume henceforth that $\lamp$ exists.  The tail
exponents satisfy the characteristic equation
\begin{subequations}\label{eq:lamLaplace}
\begin{align}
    \frac{1}{\Rn \xpm}
    &\;=\; \Lap{g}(\lampm)
    \quad\;\equiv\; \Lappm \,,
    \label{eq:lamLaplace;lam}
\intertext{where $\Lap{g}(\lambda)$ denotes the Laplace transform of
the generation interval distribution $g(\aoi)$,}
    \Lap{g}(\lambda)
    &\;=\; \int_0^\infty e^{-\lambda\aoi}g(\aoi)\,\dee\aoi \,.
    \label{eq:lamLaplace;Laplace}
\end{align}
\end{subequations}
For the epidemic trajectories considered here, $\Rn\xm>1$, while
$\Lap{g}(\lambda)$ is continuous and strictly decreasing from 1 to 0
as $\lambda$ increases over the positive half-line.  Hence the rising
characteristic equation has a unique solution $\lamm>0$.  We will
refer to the asymptotic exponential rates as the \term{tail
exponents}; formally, $\lampm$ are Lyapunov characteristic
exponents \citeref{Stro18,BarrPesi02} obtained by linearizing about
the points $(\xm,0)$ and $(\xp,0)$, which are the limits of the
trajectory $(X(\tau),Y(\tau))$ as $\tau \to \pm\infty$.

For either tail, \cref{eq:lamLaplace;lam} expresses a standard
Euler--Lotka relationship \cite{DublLotk25,Fell41,KeyfCasw05}.  If
$\xm=1$, \cref{eq:lamLaplace;lam} with a minus sign (the rising tail)
reduces to the formula relating growth rates and reproduction numbers
given in the epidemiological context by Wallinga and Lipsitch
(WL) \citeref[Equation~(2.7)]{WallLips07}.  If $\xm<1$, the rising
tail determines only $\Rn\xm=\Reff$, not $\Rn$
itself \citeref{McCaw+2009}.

The falling-tail Euler--Lotka relationship links $\lamp$, $g(\aoi)$,
and $\Rn\xp$.  Combining it with the relation between the
pre-epidemic product $\Rn\xm$ and the post-epidemic product $\Rn\xp$
supplied by the conservation law \labelcref{eq:FI} yields a
relationship between $\lamp$ and $\Rn\xm$
[\appref[rn-from-lamp-main]{app:Rn.from.lamp};
\cref{eq:Rn.via.Lapp}] that, to our knowledge, has not been recognized
previously.

For the models we have emphasized, expressions for
$g(\aoi)$, $\Lap{g}(\lambda)$, and $\lampm$
are given in \cref{tab:GI.properties}.

\begin{table}
\centering
\caption{Intrinsic generation-interval distributions, Laplace
transforms, and tail exponents for the SIR model \labelcref{eq:SIR},
SEIR model \labelcref{eq:SEIR}, and the \RE \cref{eq:re} with a Gamma
distributed intrinsic generation interval $g(\aoi)$ ($\aoi$ denotes
age of infection).  In the SEIR row, $\ell$ is the ratio of the mean
latent period to the mean infectious period.  In the Gamma GI row, $a>0$
and $b>0$ are the Gamma shape and rate parameters, respectively, so the
mean is $a/b$ and the variance is $a/b^2$.
}
\label{tab:GI.properties}
\renewcommand{\arraystretch}{1.5}  % increases row/interline spacing
\setlength{\tabcolsep}{10pt}       % increases inter-column padding
\begin{tabular}{llll}
\hline
Model
&
$g(\aoi)$
&
$\Lap{g}(\lambda)$
&
$\lampm$
\\
\hline
SIR
&
$\displaystyle e^{-\aoi}$
&
$\displaystyle \frac{1}{\lambda+1}$
&
$\displaystyle \Rn\,\xpm - 1$
\\[2ex]
SEIR
&
$\displaystyle
  \hspace{-10pt}\begin{cases}
    \aoi\,e^{-\aoi}, & \ell=1 \\[5pt]
    \dfrac{e^{-\aoi}-e^{-\aoi/\ell}}{1-\ell}, & \ell\ne1
  \end{cases}$
&
$\displaystyle
  \frac{1}{\ell\lambda+1}\cdot\frac{1}{\lambda+1}$
&
$\displaystyle
  \frac{2(\Rn\xpm-1)}
       {\sqrt{(1-\ell)^2+4\ell\Rn\xpm}+(1+\ell)}
$
\\[5ex]
Gamma GI
&
$\displaystyle
  \frac{b^{a}}{\Gamma(a)}
  \aoi^{a-1}e^{-b\aoi}$
&
$\displaystyle
  \Big(\frac{b}{\lambda+b}\Big)^{a}$
&
$\displaystyle
  b\big((\Rn\xpm)^{1/a} - 1\big)$
\\
\hline
\end{tabular}

\end{table}

\beginsubsection{Momentum, endpoints, and tail exponents}{sec:est.Rn.xm}

For each tail exponent, \cref{eq:lamLaplace;lam}
determines the corresponding effective reproduction number
at each epidemic endpoint ($\Reff = \Rn\xm$ or $\Rn\xp$), 
but not the individual factors.  Evaluating the first integral
\labelcref{eq:FI} at the endpoints $(\xpm,0)$ and at the momentum peak
$(\xpeak,\ypeak)$ gives $\Rn\ypeak=\Volterra(\Rn\xpm)$; for either
exponent, \cref{eq:lamLaplace;lam} therefore gives
\begin{equation}\label{eq:R.C.lampm}
\Rn\ypeak
\;=\;
\Volterra(\Rn\xpm)
\;=\;
\Volterra\big(\frac{1}{\Lappm}\big) \,.
\end{equation}
Combining \cref{eq:lamLaplace;lam,eq:zm,eq:zp} gives the
endpoint identities
\begin{subequations}\label{eq:compute}
\begin{align}
\zm &\;=\; 1 - \frac{1}{\Rn\Lapm} \,, \label{eq:zm.compute} \\
\zp &\;=\; \frac{1}{\Rn}\Big(\frac{1}{\Lapm} - \frac{1}{\Lapp}\Big) \,. \label{eq:zp.compute}
\end{align}
\end{subequations}
The two endpoint forms of \cref{eq:R.C.lampm} imply
\begin{equation}\label{eq:V=V}
\Volterra\big(\frac{1}{\Lapm}\big)
\;=\;
\Volterra\big(\frac{1}{\Lapp}\big) \,.
\end{equation}
Thus, the two tail exponents are
constrained by \cref{eq:V=V}; given either exponent and $g$, the other
is the unique solution on the opposite side of zero.
Identities \labelcref{eq:compute,eq:V=V} provide the mathematical basis for joint
inference of transmissibility and prior immunity from epidemic time
series \citeref{EarnPars26_epimom_inference}.

\beginsection{Discussion}{sec:Discussion}

We have identified the epidemic momentum $Y(\tau)$ [\cref{eq:re.Y}] as
a quantity of fundamental interest for analysis of infectious disease
dynamics.  In the simple SIR and SEIR models, epidemic momentum
corresponds to prevalence of infection, but in general it provides
different information than prevalence: rather than simply counting
the number of infected individuals, the epidemic momentum weights
individuals by their potential to infect others in the future, which is
more predictive of the future spread of disease.

We have shown very generally that epidemic models possess a first
integral (a quantity that is conserved along epidemic trajectories),
which is a simple function of the susceptible proportion of the
population and the epidemic momentum [\cref{eq:FI}], the fixed value
of which is the peak epidemic momentum $\ypeak$ [\cref{eq:FI.value}].
The explicit expression for the conserved quantity \labelcref{eq:FI}
for a generic epidemic yields an exact solution in the
susceptible-momentum ($x$-$y$) phase plane [\cref{fig:phaseplane}],
with a universal functional form \labelcref{eq:Yofx} that is identical
for any model that can be expressed using the renewal equation
\labelcref{eq:re}.  All that varies among models is the speed with
which trajectories in the $x$-$y$ phase plane are traversed
[\cref{eq:timechange}].  Thus, identifying the momentum has revealed a
broad geometric invariance of epidemics (which generalizes to models
with nonlinear incidence \citeref{epimom-nli}).
The invariance of the full phase plane geometry has
valuable theoretical and practical implications.

The initial growth rate and generation interval distribution $g(\aoi)$
determine only the initial effective reproduction number,
$\Reff=\Rn\xm$, where $\xm$ is the population proportion that was
susceptible before the outbreak began
\citeref{WallLips07,McCaw+2009}.  Knowing this product does not
distinguish pathogen transmissibility from prior susceptibility.  The
incidence history and $g(\aoi)$ then determine epidemic momentum and
$\Reff(\tau)$ through time.  Evaluating the conservation law
\labelcref{eq:FI} using these quantities supplies the additional
relation needed to determine $\Rn$ and $\xm$ separately.  This result
provides the mathematical basis for jointly inferring $\Rn$ and $\xm$
from an epidemic time series
\citeref{EarnPars26_epimom_inference}.

An immediate and essentially trivial consequence of the universality
of the phase plane dynamics is that the expressions for the pre- and
post-epidemic susceptible proportions [\cref{eq:xpm}] are equally
universal.  Classical final-size relations express the asymptotic
post-epidemic susceptible fraction in terms of an epidemic state
specified at a finite initial
time \citeref{KermMcKe27,MaEarn06,Mill2012}.  Our complete-trajectory
analysis instead determines both $\xm$ and $\xp$ from $\Rn$ and any
specified state on the trajectory [\cref{eq:xpm}], so prior population
immunity need not be supplied
\emph{a priori}.  It thereby yields the \emph{genuine} final size of
the focal epidemic, $\zp=\xm-\xp$ [\cref{eq:zp}], and distinguishes it
from the total post-epidemic immune fraction, $\zptot=1-\xp$
[\cf \cref{eq:zp.vs.naive}].  
Our derivation reveals that the final-size relation is model-invariant
because it is determined by the endpoints of model-invariant complete
trajectories in the susceptible-momentum phase plane.

\beginsubsection{Physics analogies}{sec:physics.analogies}

We chose the terminology ``epidemic momentum'' after considering the
meanings of the components of the generic equation
\begin{equation}\label{eq:dY.again}
\ddtau{Y}=(X-\xpeak)\FoI,
\end{equation}
which is equivalent to \cref{eq:SIR;Y,eq:XYF;Y,eq:re.Ydot} in
the specific cases that we have discussed.

While the factor $\FoI$ in \cref{eq:dY.again} is traditionally called
the ``force of infection'' (\FoItext), it does not behave like a
physical force, which would vanish in the absence of something to act
on (\eg in Newton's Second Law, $\mathbf{F}=m\mathbf{a}$, the force
vanishes if the mass is zero).  The \FoItext determines the hazard
that any given susceptible individual faces---and might cause them to
become infected---regardless of how many other susceptible individuals
there are.  It is natural, therefore, to think of the \FoItext as an
``infective field'' or, to avoid changing the acronym, a \term{field
of infectivity} that induces a force on any susceptible individuals.

If the factor $X-\xpeak$ in \cref{eq:dY.again} were simply $X$, then
we might reasonably think of $X$ as analogous to mass and $\FoI$ as
analogous to a gravitational field.  \cref{eq:dY.again} would then be
analogous to Newton's Second Law, with $X$ playing the role of $m$ and
$\FoI$ playing the role of $\mathbf{a}$.  Since $m=X$ is not constant
in our context, we should instead use the more general form of
Newton's law, namely $\mathbf{F} = \ddt{\mathbf{p}}$, where
$\mathbf{p}=m\mathbf{v}$ is the momentum.  This loosely motivates
us to think of $Y$ as a momentum in \cref{eq:dY.again}.

However, $\xpeak>0$ in our systems, so $X-\xpeak\ne X$ and we do not
yet have a direct analogy with a physical system.  We can construct a
more convincing analogy by attributing a \emph{charge} to each
individual in our population.  To that end, we imagine
three particle types, analogous to protons,
electrons, and neutrons.  Unlike the usual $1,-1,0$ charge convention
in electromagnetism, however, two of the three types of ``epidemic particles''
have the same charge: we assign charges $1-\xpeak$, $-\xpeak$, and
$-\xpeak$ to susceptible, infected (momentum-bearing), and removed
individuals, respectively.  If
$Z=1-X-Y$ denotes the removed fraction in this bookkeeping, the net
population charge density or \term{epidemic charge} is
\begin{equation}\label{eq:epidemic.charge}
  Q \;=\; X(1-\xpeak) + Y(-\xpeak) + Z(-\xpeak) \;=\; X-\xpeak,
\end{equation}
so the \term{epidemic force} is $Q\FoI$.  The charge convention is
chosen so that zero charge corresponds to the state $X=\xpeak$, \ie
the ``neutral state'' in which momentum is at its peak and the force
vanishes.  We have described this situation in language that suits the
SIR model, but \cref{eq:epidemic.charge} remains valid in the
generality of the \RE, for which the natural definition of $Z$ (like
$Y$) is a weighted integral over incidence history and $X+Y+Z=1$ still
holds; see \appref[initial-conditions-charge]{app:general.z}.

While we do not use physics analogies in this paper beyond motivating
the term ``epidemic momentum'', it is possible that such concepts may
prove fruitful in the future.  For readers who are interested in
connections with physics, it is worth mentioning that in the special
case of the SIR model, there is an exact structural correspondence
with mechanics.  The SIR model is a standard Hamiltonian
system \citeref{Arno13} with canonical coordinate $\ln{X}$ and
conjugate momentum $\ln{Y}$ (the conjugate momentum is the logarithm
of the epidemic momentum $Y$, \emph{not} $Y$ itself).  The standard
SIR equations follow from the Hamiltonian
${\mathcal{H}}(\ln{X},\ln{Y})=-\Rn e^{\ln{X}}+\ln{X}-\Rn e^{\ln{Y}}$,
and this Hamiltonian structure is retained if $\Rn$ is time-dependent.
The SIR model can also be viewed as a special case of the
Lotka--Volterra predator--prey system, with prey and predator
densities given by $X$ and $Y$; the full Lotka--Volterra system is
Hamiltonian in logarithmic coordinates \citeref{Stro18}.  Since the
renewal equation \labelcref{eq:re} is equivalent to the SIR model
after a change of time variable [\cref{eq:timechange}], a generic
epidemic model not only shares the SIR phase plane geometry but is
``Hamiltonian up to a change of time variable''.

\beginsubsection{Extensions and Generalizations}{sec:ext.gen}

\beginsubsubsection{Nonlinear incidence}{sec:nli}

The most common epidemic models are based on the principle of mass
action, which amounts to assuming that the population is homogeneously
mixed with contacts among hosts occurring in direct analogy with
collisions of particles in an ideal gas. In an effort to understand
the effects of heterogeneous contact structures, a substantial amount
of research has been devoted to the analysis and use of nonlinear
incidence models that attempt to mimic contact heterogeneities without
explicitly keeping track of individuals of different types
%%\citeref{WilsWorc45,WilsWorc45b,Liu+86,Liu+87,FinkGren00,Koro06,Koro07,Beck17,Novo08}.
\citeref{WilsWorc45,Liu+86,FinkGren00,Novo08}.
In most of these analyses, the incidence is taken to be nonlinear in
$X$ but still proportional to $\FoI$ (or to a function of $\FoI$), in
which case one can still easily define epidemic momentum as a weighted
integral over incidence history, derive an exact
phase-plane solution, a first integral, \etc\  We present
details elsewhere \citeref{epimom-nli}.

\beginsubsubsection{Approximation of solutions of epidemic models}{sec:approx}

As mentioned in the introduction, \KM built on their phase plane
solution to find a (local) approximation to the SIR temporal
dynamics \cite[p.\,714]{KermMcKe27}.  We show elsewhere \citeref{gaem}
that an extremely accurate, globally valid, analytical approximation
for the epidemic momentum $Y(\tau)$ can be derived generically, from
which we derive accurate analytical approximations to the \FoItext and
incidence.

\beginsubsubsection{Time-dependent transmission rates}{sec:time.dep}

When a new disease emerges, the transmission rate ($\beta$) inevitably
changes as a result of changes in human behaviour, either imposed by
policies such as lockdowns or school
closures \citeref{Cauc+08,Earn+12}, or as a result of fear or
caution \citeref{BootFerg07,He+13a}.  Changes in transmission rate,
either resulting from such exogenous factors or from intrinsic changes
in transmissibility of the pathogen (\eg resulting from the emergence
of new variants), can be modelled by a time-varying $\beta$.  Again,
under commonly used assumptions, one can define epidemic momentum as an
integral over incidence history and derive \cref{eq:dY.again}.
However, unlike \cref{eq:dYdx}, the resulting phase plane equation is
non-autonomous so does not yield simple phase plane geometry
(see \appref[epimom-rt-time-dependence]{app:epimom.Rt}).

\beginsubsubsection{Susceptible recruitment}{sec:srpb}

The generic framework we have considered [\cref{eq:XYF}]
ignores sources of new susceptibles, \eg from births, immigration,
and/or decay of immunity.  However, in the presence of vital dynamics
(births and deaths) and other forms of susceptible recruitment,
epidemic momentum is still meaningful and defined in exactly the same
way, via \cref{eq:re.Y} (or the equivalents under nonlinear
incidence \citeref{epimom-nli}).

\beginsubsubsection{Population momentum more generally}{sec:pop.mom}

The concept of epidemic momentum arose from our analysis of epidemic
models, but it seems likely that the notion of a \term{population
momentum} may lead to fruitful developments in other areas of
population dynamics \citeref{BrauCast12}.  More generally, dynamical
models in other areas of biology \citeref{Efti+11}, other
sciences \citeref{simon2020sirchemical}, and the social
sciences \citeref{Bass1969}, often have structure that resembles
epidemic models, and may benefit from analyses similar to those we
have introduced here.

% Keep the short Conclusion together before the arXiv declaration page.
\ifepimomarxiv
  \enlargethispage{2\baselineskip}
\fi
\beginsubsection{Conclusion}{sec:conclusion}

Identifying epidemic momentum as a state variable exposes a universal
phase portrait for epidemics that are governed by renewal equations.
Differences in generation interval distributions affect the rate at
which trajectories are traversed, but not their form in the phase
plane. Euler-Lotka relationships connect the rising and falling tail
exponents to the corresponding epidemic endpoints.  From $\Rn$ and any
one state on a trajectory in the susceptible-momentum phase plane, a
conservation law determines both endpoint susceptible fractions, and
hence prior population immunity and the genuine epidemic final size.
Thus, fundamental epidemiological quantities are determined by a
single exact geometry across the entire class of \RE models.

%%\typeout{get arXiv to do 4 passes: Label(s) may have changed. Rerun}

\clearpage

\section*{Declaration of AI use}
\msaiuse

\ifanonymous\else
  \section*{Funding}
  \msfunding

  \section*{Acknowledgements}
  \msacknowledgements
\fi

\appendix
%% ensure equations are numbered by section: A1, A2, ... B1, B2, ...
\numberwithin{equation}{section}
\renewcommand{\theequation}{\thesection\arabic{equation}}

\section*{APPENDICES}
%%\addcontentsline{toc}{section}{Appendix A}
%%\stepcounter{section}

%% Appendix driver for the epimom theory manuscript.
%% Appendices cited from the main text or in appendices with a citation path that ends in the main text, ordered by first citation assuming the reader follows forward citation paths to appendices.

\beginappendix{Equivalence of the \RE and compartmental models}{app:re.to.seir}
\appfilename{epimom_theory/epimom_app_re_to_seir.tex}
\appbackrefs{\appbackref{re-to-seir-gi}{generation-interval discussion};
\appbackref{re-to-seir-momentum}{epidemic momentum for SIR and SEIR}}

The standard (dimensionless) generation interval distributions for the
SIR and SEIR models are listed in \cref{tab:GI.properties}.  In
Ref.~\citeref{Cham+18}, these distributions are derived from the
standard ODEs [\cref{eq:SIR,eq:SEIR} in this paper].  Here we do the
reverse: we start from the \RE \labelcref{eq:re} with the putative
generation interval distribution $g(\aoi)$ and derive the ODEs for the
SIR and SEIR models.  This calculation confirms that the epidemic
momentum $Y$ reduces to total prevalence for these simple models.

In the SEIR model \labelcref{eq:SEIR}, the \FoItext is
$\FoI=\Rn\Iprop$, so incidence is $\inc=\Rn X\Iprop$ and the
\RE \labelcref{eq:re} can be written
\begin{subequations}
\begin{align}
   \ddtau{X} &\;=\; -\inc(\tau), \label{eq:re.SEIR;X}\\
   \Iprop(\tau) &\;=\; \int_{0}^{\infty} \inc(\tau-\aoi)g(\aoi)\, \dee \aoi. \label{eq:re.SEIR;Iprop}
\end{align}
\end{subequations}
In the SIR case ($\ell\to0$ in \cref{tab:GI.properties}), $g(\aoi) =
e^{-\aoi}$, so the reduced reproduction
number \labelcref{eq:Ra} is $\Ra=\Rn g(\aoi)$;
hence \cref{eq:re.SEIR;Iprop} implies that $\Iprop$ is the epidemic
momentum $Y$ \labelcref{eq:re.Y} for the SIR model.

Differentiating \cref{eq:re.SEIR;Iprop} and integrating by parts, we find
\begin{subequations}
\begin{align}
   \ddtau{\Iprop} 
   &\;=\; \int_{0}^{\infty} \inc'(\tau-\aoi)g(\aoi)\, \dee \aoi \nonumber \\
   %%&\;=\; -\inc(\tau-\aoi)g(\aoi)\big|_{0}^{\infty} + \int_{0}^{\infty} \inc(\tau-\aoi)g'(\aoi)\, \dee \aoi\\
   &\;=\; \inc(\tau)g(0) + \int_{0}^{\infty} \inc(\tau-\aoi)g'(\aoi)\, \dee \aoi. \label{eq:ddtauIprop}
\end{align}
\end{subequations}
In the SIR case, $g(\aoi) = e^{-\aoi}$ implies $g(0) = 1$ and $g'(\aoi) = -g(\aoi)$, whence
\begin{equation}
   \ddtau{\Iprop} \;=\; \inc(\tau) - \int_{0}^{\infty} \inc(\tau-\aoi)g(\aoi)\, \dee \aoi \;=\; \Rn X\Iprop - \Iprop,
\end{equation}
which is an ODE that is identical to \cref{eq:SIR;Y} since $Y=\Iprop$ for the SIR model.
  
In the SEIR case, differentiating $g(\aoi)$ [\cref{tab:GI.properties}] yields
\begin{equation}
   g'(\aoi) \;=\;
    \begin{cases}
       (1 - \aoi)e^{- \aoi}
            & \quad \ell\;=\;1, \\[5pt]
       \frac{1}{1-\ell}\big(\frac{1}{\ell} e^{-\frac{\aoi}{\ell}}- e^{- \aoi}\big)
            & \quad \ell \;\ne\; 1.
    \end{cases}
\end{equation}
Focusing on the generic case ($\ell\ne 1$), we note that $g(0)=0$ and
$g'(\aoi) = \frac{1}{\ell} e^{-\frac{\aoi}{\ell}} -  g(\aoi)$,
%% \begin{subequations}
%% \begin{align}
%% 	g'(\aoi) 
%% %	&\;=\;\frac{1}{1-\ell}\Big(\frac{1}{\ell} e^{-\frac{\aoi}{\ell}}- e^{- \aoi}\Big)\\
%% 	&\;=\;\frac{1}{1-\ell}\Big(\big(\frac{1}{\ell}-1\big) e^{-\frac{\aoi}{\ell}}- \big(e^{- \aoi}-e^{-\frac{\aoi}{\ell}}\big)\Big)\\
%% 	&\;=\;  \frac{1}{\ell} e^{-\frac{\aoi}{\ell}} -  g(\aoi),
%% \end{align}
%% \end{subequations}
so \cref{eq:ddtauIprop} becomes
\begin{equation}
   \ddtau{\Iprop} \;=\; \frac{1}{\ell} \int_{0}^{\infty} \inc(\tau-\aoi) e^{-\frac{\aoi}{\ell}}\, \dee \aoi -  \Iprop(\tau).
\end{equation}
If we now differentiate the integral on the right hand side, and then integrate by parts, we find
\begin{subequations}
\begin{align}
   \ddtau{} \int_{0}^{\infty} \inc(\tau-\aoi) e^{-\frac{\aoi}{\ell}}\, \dee \aoi 
   &\;=\;  \int_{0}^{\infty} \inc'(\tau-\aoi) e^{-\frac{\aoi}{\ell}}\, \dee \aoi\\
   %%&\;=\;  -\inc(\tau-\aoi) e^{-\frac{\aoi}{\ell}} \big|_{0}^{\infty} - \frac{1}{\ell} \int_{0}^{\infty} \inc(\tau-\aoi) e^{-\frac{\aoi}{\ell}}\, \dee \aoi\\
   &\;=\; \Rn X(\tau)\Iprop(\tau) - \frac{1}{\ell} \int_{0}^{\infty} \inc(\tau-\aoi) e^{-\frac{\aoi}{\ell}}\, \dee \aoi,
\end{align}
\end{subequations}
where we have used $\inc(\tau) = \Rn X(\tau)\Iprop(\tau)$.  Thus, if we define
\begin{equation}\label{eq:SEIR-E}
   \Eprop(\tau) \;=\;  \int_{0}^{\infty} \inc(\tau-\aoi) e^{-\frac{\aoi}{\ell}}\, \dee \aoi
\end{equation}
then we recover the SEIR compartmental equations \labelcref{eq:SEIR}.
We can see that this definition of $\Eprop$ is correct biologically
by noting that if
$\Tlat$ is the random variable giving the length of the
latent period then
\begin{equation}
	 e^{-\frac{\tau}{\ell}} \;=\; {\mathbb P}\{\Tlat > \tau\},
\end{equation}
so the integral in \cref{eq:SEIR-E} counts individuals who were
infected at some time in the past and are not yet infectious at time
$\tau$.  If we now consider the sum of the exposed and infectious
variables, we find
\begin{subequations}
\begin{align}
\Eprop(\tau) + \Iprop(\tau) &\;=\; \int_{0}^{\infty} \inc(\tau-\aoi) \big(g(\aoi) + e^{-\frac{\aoi}{\ell}}\big)\, \dee \aoi\\
   %%&= \int_{0}^{\infty} \inc(\tau-\aoi) \big(\frac{e^{-\aoi} - e^{-\aoi/\ell}}{1 - \ell} + e^{-\frac{\aoi}{\ell}}\big)\, \dee \aoi\\
   &\;=\; \int_{0}^{\infty} \inc(\tau-\aoi) \big(\frac{e^{-\aoi} - \ell e^{-\aoi/\ell}}{1 - \ell}\big)\, \dee \aoi\\
   &\;=\;  \int_{0}^{\infty} \inc(\tau-\aoi) \int_{\aoi}^{\infty} g(\aoidum)\, \dee \aoidum\, \dee \aoi,
\end{align}
\end{subequations}
so, again, referring to \cref{eq:GI.survival,eq:re.Y}, we see that for the SEIR
model \labelcref{eq:SEIR}, $\Eprop+\Iprop$ coincides with epidemic
momentum $Y$ \labelcref{eq:re.Y}.
 % app:re.to.seir
\beginappendix
    {\texorpdfstring{Equivalence of integral and ODE representations of epidemic momentum $Y(\tau)$}{Equivalent representations of epidemic momentum Y(tau)}}{app:y.equiv}
\appfilename{epimom_theory/epimom_app_y_equiv.tex}
\appbackrefs{%
  \appbackref{yinit-from-re}{$\yinit$ from \RE};
  \appbackref{proof-y-cuminc}{cumulative-incidence representation};
  \appbackref{proof-y-recover}{recovering force of infection and incidence};
  \appbackref{proof-y-david-equivalence}{integral--ODE equivalence note};
  \appbackref{epimom-rt-y-to-re}{Appendix on time-varying reproduction number: differentiating the integral form}
}

\beginsubappendix{The \RE and the ODE for $Y$ imply the integral form of $Y$}{subapp:re.to.Y}

Using \cref{eq:re;FoI,eq:re;inc} and setting $\xpeak = \frac{1}{\Rn}$, we can re-write \cref{eq:XYF;Y} as
\begin{align}\label{eq:Y.ode.inc}
	\ddtau{Y} &\;=\; \inc(\tau) - \xpeak \FoI(\tau)
	\quad\;=\; \inc(\tau) - \int_{0}^{\infty} \inc(\tau-\aoi) g(\aoi)\, \dee \aoi.
\end{align}
Integrating left and right hand sides, and multiplying the first term on the right by $\int_0^\infty g(\aoi)\,\dee\aoi=1$, we have
\begin{subequations}\label{eq:Y.int.inc}
\begin{align}
Y(\tau) - \yinit 
	&\;=\; \int_0^\infty g(\aoi)\,\dee\aoi \int_{\tauinit}^{\tau} \inc(\taudum)\, \dee \taudum
	- \int_{0}^{\infty} \int_{\tauinit}^{\tau} \inc(\taudum-\aoi)\, \dee \taudum\, g(\aoi)\, \dee \aoi
        \\
	&\;=\; \int_{0}^{\infty}
	\left(
	\int_{\tauinit}^{\tau} \inc(\taudum)\, \dee \taudum
	-
	\int_{\tauinit}^{\tau} \inc(\taudum-\aoi)\, \dee \taudum
	\right)
	g(\aoi)\, \dee \aoi \\
	&\;=\; \int_{0}^{\infty}
	\left(
	\int_{\tauinit}^{\tau} \inc(\taudum)\, \dee \taudum
	-
	\int_{\tauinit-\aoi}^{\tau-\aoi} \inc(\taudum)\, \dee \taudum
	\right)
	g(\aoi)\, \dee \aoi \\
%% \intertext{so \ $
%% \int_{\tauinit}^{\tau} - \int_{\tauinit-\aoi}^{\tau-\aoi}
%% = \int_{\tauinit}^{\tau} - \left( \int_{\tauinit-\aoi}^{\tauinit} + \int_{\tauinit}^{\tau-\aoi} \right)
%% = \int_{\tauinit}^{\tau} - \left( \int_{\tauinit-\aoi}^{\tauinit} - \int_{\tau-\aoi}^{\tauinit} \right)
%% = \int_{\tau-\aoi}^{\tauinit} + \int_{\tauinit}^{\tau} - \int_{\tauinit-\aoi}^{\tauinit}
%% = \int_{\tau-\aoi}^{\tau} - \int_{\tauinit-\aoi}^{\tauinit}
%% $ \ yields}
	&\;=\; \int_{0}^{\infty}
	\left(
	\int_{\tau-\aoi}^{\tau} \inc(\taudum)\, \dee \taudum
	-
	\int_{\tauinit-\aoi}^{\tauinit} \inc(\taudum)\, \dee \taudum
	\right)
	g(\aoi)\, \dee \aoi. \label{eq:integral-diff}
\end{align}
\end{subequations}
Now consider the first inner integral, from $\tau-\aoi$ to $\tau$,
in which $\aoi$ is constant, and make the change of variables
\;$\taudum \;=\; \tau - \aoidum,$\;
so
$\dee \taudum = -\dee \aoidum$.  When
$\taudum = \tau - \aoi$, we have $\aoidum = \aoi$, and when
$\taudum = \tau$, we have $\aoidum = 0$.  Hence
\begin{align}
\int_{\tau-\aoi}^{\tau} \inc(\taudum)\,\dee \taudum
&\;=\;
\int_{\aoi}^{0} \inc(\tau-\aoidum)(-\dee \aoidum) 
\;=\;
\int_{0}^{\aoi} \inc(\tau-\aoidum)\,\dee \aoidum .
\label{eq:inner-rewritten}
\end{align}
Substituting \cref{eq:inner-rewritten} into the first term of \cref{eq:integral-diff}
gives
\begin{equation}
\label{eq:double-integral-before-switch}
\int_{0}^{\infty}
  \left(
  \int_{0}^{\aoi} \inc(\tau-\aoidum)\,\dee \aoidum
  \right)
  g(\aoi)\,\dee \aoi ,
\end{equation}
which is an integral over the triangular region
$0 \leq \aoidum \leq \aoi < \infty$.
If we reverse the order of integration, the same region is described by
$0 \leq \aoidum < \infty$ and $\aoi \geq \aoidum$.
Therefore, assuming the usual hypotheses needed to justify Fubini's theorem,
\begin{align}
\int_{0}^{\infty}
  &\left(
    \int_{0}^{\aoi} \inc(\tau-\aoidum)\,\dee \aoidum
  \right)
  g(\aoi)\,\dee \aoi 
\;=\;
\int_{0}^{\infty}
  \int_{\aoidum}^{\infty}
  \inc(\tau-\aoidum) g(\aoi)
  \,\dee \aoi\,\dee \aoidum  \nonumber \\
&\;=\;
\int_{0}^{\infty}
  \inc(\tau-\aoidum)
  \left(
  \int_{\aoidum}^{\infty} g(\aoi)\,\dee \aoi
  \right)
  \dee \aoidum 
\;=\;
\int_{0}^{\infty}
  \inc(\tau-\aoi)
  \left(
  \int_{\aoi}^{\infty} g(\aoidum)\,\dee \aoidum
  \right)
  \dee \aoi ,
\label{eq:first-final}
\end{align}
where in
the last step we have simply swapped the names
of the dummy variables
$\aoidum$ and $\aoi$.

The second inner integral in \cref{eq:integral-diff} is identical
in form to the first, so we finally have that the full \cref{eq:integral-diff}
can be written
\begin{equation}
\int_{0}^{\infty} \inc(\tau-\aoi) \left(\int_{\aoi}^{\infty} g(\aoidum)\, \dee \aoidum\right)\dee \aoi \nonumber
	- \int_{0}^{\infty} \inc(\tauinit-\aoi) \left(\int_{\aoi}^{\infty} g(\aoidum)\, \dee \aoidum\right)\dee \aoi \,.
\end{equation}
Using \cref{eq:GI.survival} to simplify, and inserting in \cref{eq:Y.int.inc}, we have
\begin{equation}
Y(\tau) - \yinit 
\;\ = \;\ 
\int_{0}^{\infty} \inc(\tau-\aoi)\gisurv(\aoi)\, \dee \aoi
         \ - \ \int_{0}^{\infty} \inc(\tauinit-\aoi)\gisurv(\aoi)\, \dee \aoi \,, \label{eq:Y.int.inc;d}
\end{equation}
yielding \cref{eq:re.Y}, with the self-consistent initial condition stated in \cref{eq:re;yinit}.

\beginsubappendix{The integral form and ODE for $Y$ together imply the \RE}{subapp:Y.to.re}

In \appref{subapp:re.to.Y}, using \cref{eq:XYF;Y}, we showed that
if $\FoI(\tau)$ is obtained via the \RE \labelcref{eq:re}, then the
integral representation \labelcref{eq:re.Y} of $Y(\tau)$ is implied.
Conversely, we show here that the integral representation \cref{eq:re.Y} for $Y(\tau)$, together
with the ODE \cref{eq:XYF}, yields the \RE \labelcref{eq:re}.

First, \cref{eq:re;X,eq:XYF;X} for $\ddtau{X}$ are identical.
Next, inserting \cref{eq:re.Y} into \cref{eq:XYF;Y}, and integrating by parts, we have
\begin{equation}\label{eq:FoI.via.int.by.parts}
  (X(\tau) - \xpeak)\FoI(\tau) \;=\; \ddtau{Y}  \;=\; \inc(\tau) - \int_{0}^{\infty} \inc(\tau-\aoidum) g(\aoidum) \, \dee \aoidum.
\end{equation}
Recalling the definition \labelcref{eq:re;inc} of $\inc(\tau)$, \cref{eq:FoI.via.int.by.parts} is equivalent to
\begin{equation}
	(X(\tau) - \xpeak)\FoI(\tau) \;=\; X(\tau)\FoI(\tau) - \int_{0}^{\infty} X(\tau-\aoidum) \FoI(\tau-\aoidum) g(\aoidum) \, \dee \aoidum.
\end{equation}
Subtracting the common term $X(\tau)\FoI(\tau)$ from both sides
and multiplying both sides by $-\Rn$
yields \cref{eq:re;FoI}.
 % app:y.equiv
\beginappendix{\texorpdfstring{Asymptotic growth rates $\lampm$ from the renewal equation}{Asymptotic growth rates lambda+/- from the renewal equation}}{app:lampm.re}
\appfilename{epimom_theory/epimom_app_lampm_re.tex}
\appbackrefs{\appbackref{yinit-tail-exponents}{Appendix on initial epidemic momentum: initial exponential growth};
\appbackref{tail-exponents-main}{main text: tail-exponent derivation};
\appbackref{epimom-rt-tail-exponents}{Appendix on time-varying reproduction number: tail exponents}}

\beginsubappendix{$\lampm$ for the SIR model}{subapp:lampm.re.sir}

For the specific example of the SIR model, $\FoI=\Rn Y$,
so \cref{eq:XYF;Y} implies that
\begin{equation}
\ddtau{\,\log{\FoI}} \;=\; \ddtau{\,\log{Y}} \;=\; \Rn(X - \xpeak).
\end{equation}
Thus, $\FoI$ and $Y$ have the
same exponential growth rates at all times.  For early and late times
($\tau\to\pm\infty$), the susceptible fraction $X(\tau)\to\xpm$
[\cref{eq:xpm}], so
\begin{equation}
\Rn(X - \xpeak) \to \Rn(\xpm - \xpeak).
\end{equation}
Moreover, $\ddtau{\inc}\to \Rn \xpm \ddtau{Y}$, so
\begin{equation}
\ddtau{\,\log{\inc}} \to \ddtau{\,\log{Y}}.
\end{equation}
Thus, $\FoI$, $Y$, and
$\inc$ have identical exponential rates of change asymptotically.
In addition, since $\xm> \xpeak > \xp$ [\cref{eq:xp.lt.xm}], it
follows that it is exponential \emph{growth} as $\tau\to-\infty$ and
exponential \emph{decay} as $\tau\to+\infty$.

\beginsubappendix{$\lampm$ from the \RE}{subapp:lampm.re.generic}

More generally, consider either asymptotic limit of a solution
of the \RE \labelcref{eq:re}.  As $\tau\to\pm\infty$,
$X(\tau)\to\xpm$ [\cref{eq:xpm}].  Linearizing
\cref{eq:re;FoI} about the corresponding endpoint gives a homogeneous
Lotka renewal equation
\citeref[Chapter~20, pp.\,353--364]{Kot2001}.  We first consider
solutions for which the \FoItext is asymptotically exponential in the
corresponding limit,
$\FoI(\tau)\sim\FoIpm e^{\rpm\tau}$, where
$\Lap{g}(\rpm)<\infty$.  The existence of a corresponding tail
exponent is addressed in \appref{app:lampm.re.exist}.  It follows that
\begin{equation}\label{eq:inc;tail}
	\inc(\tau) \;=\; X(\tau)\FoI(\tau) \sim \xpm\FoIpm e^{\rpm \tau}.
\end{equation}	
Inserting these asymptotic expressions into
\cref{eq:re;FoI}, we have
  \begin{align}
    \FoIpm e^{\rpm\tau}\;=\; \Rn \xpm \FoIpm e^{\rpm\tau}
    \int_0^\infty e^{-\rpm \aoi} g(\aoi)\,\dee \aoi
  \end{align}
and hence
  \begin{equation}\label{eq:rLaplace}
    \frac{1}{\Rn \xpm} \;=\; \int_0^\infty e^{-\rpm \aoi} g(\aoi)\,\dee \aoi \;=\; \Lap{g}(\rpm),
  \end{equation}
where $\Lap{g}$ denotes the Laplace transform of $g(\aoi)$ as in \cref{eq:lamLaplace;Laplace}.

If we now insert \cref{eq:inc;tail} in \cref{eq:re.Y}, we obtain
\begin{align}
	Y(\tau)
	%%\;=\; \int_{0}^{\infty} \inc(\tau-\aoi) \gisurv(\aoi) \, \dee \aoi
	&\;\sim\; \int_{0}^{\infty} \xpm \FoIpm e^{\rpm(\tau-\aoi)}
	\gisurv(\aoi) \, \dee \aoi
	\;=\; \xpm \FoIpm e^{\rpm\tau}  \int_{0}^{\infty} e^{-\rpm \aoi}
	\gisurv(\aoi) \, \dee \aoi,
\intertext{so the assumed asymptotic form
$\FoI(\tau)\sim\FoIpm e^{\rpm\tau}$ implies that $Y(\tau)$ has the same
asymptotic exponential rate [\cref{eq:lamLaplace;lam}], \ie
$\lampm=\rpm$.  Using \cref{eq:inc;tail} again, for
$\tau\to\pm\infty$ we have}
	Y(\tau) &\;\sim\; \inc(\tau) \int_{0}^{\infty} e^{-\lampm \aoi}
	\gisurv(\aoi) \, \dee \aoi,
        %%\quad\;\equiv\; \inc(\tau)K,
        \label{eq:Yinc}
\intertext{where the constant of proportionality between the asymptotic forms of $Y(\tau)$ and $\inc(\tau)$ is}
	\int_{0}^{\infty} e^{-\lampm \aoi}
	\gisurv(\aoi) \, \dee \aoi
        %%K
	&\;=\; \int_{0}^{\infty} e^{-\lampm \aoi} \int_{\aoi}^{\infty} g(\aoidum)\, \dee \aoidum \, \dee \aoi
	 \;=\; \int_{0}^{\infty}  \int_{0}^{\aoidum} e^{-\lampm \aoi}\, \dee \aoi \, g(\aoidum)\, \dee \aoidum \nonumber\\
	&\;=\;  \int_{0}^{\infty} \frac{1-e^{-\lampm \aoidum}}{\lampm}\, g(\aoidum)\, \dee \aoidum
	\;=\; \frac{1-\Lap{g}(\lampm)}{\lampm}.
        \label{eq:Yinc.const}
\end{align}
	
\beginsubappendix{Existence of the tail exponents}{app:lampm.re.exist}
\appfilename{epimom_theory/epimom_app_lampm_re.tex}

Because $g(\aoi)$ is a probability density on nonnegative infection
ages, if $r_1<r_2$ and $\Lap{g}(r_1)<\infty$, then
$e^{-r_1\aoi}>e^{-r_2\aoi}$ for every $\aoi>0$, and integration
against $g(\aoi)$ shows that
$\Lap{g}(r_1)>\Lap{g}(r_2)$.  Thus $\Lap{g}$ is strictly decreasing
wherever it is finite.  For $r\in[r_1,r_2]$, the integrand
$e^{-r\aoi}g(\aoi)$ is bounded by the integrable function
$e^{-r_1\aoi}g(\aoi)$, so Lebesgue's dominated convergence theorem
\citeref[\S11.32]{Rudin1976} shows that $\Lap{g}$ is continuous on its
domain of finiteness.  Moreover, $\Lap{g}(0)=1$ and
$\Lap{g}(r)\to0$ as $r\to\infty$.  Since $\Rn\xm>1$,
\cref{eq:rLaplace} therefore has a unique positive root $r^{-}$.
Standard asymptotic theory for the Lotka renewal equation obtained by
linearizing about the pre-epidemic endpoint then gives
$\FoI(\tau)\sim\FoI^{\smallminus} e^{r^{-}\tau}$ as
$\tau\to-\infty$
\citeref{Fell41}; see also
\citeref[Chapter~20, pp.\,353--364]{Kot2001}.  Thus the
rising tail of the nontrivial epidemic trajectory is asymptotically
exponential, with $\lamm=r^{-}$.

For the falling tail, $\Rn\xp<1$, so the characteristic equation is
$\Lap{g}(r^{+})=1/(\Rn\xp)>1$, with $r^{+}<0$.  Suppose first that
$g(\aoi)=0$ for $\aoi>A$, where $A<\infty$.  Because $g$ is a
probability density, there is some $\epsilon\in(0,A)$ for which
$\int_{\epsilon}^{A}g(\aoi)\,\dee\aoi>0$.  For every $c>0$,
\begin{equation}\label{eq:bounded.support.Laplace}
\Lap{g}(-c)
\;=\;
\int_0^A e^{c\aoi}g(\aoi)\,\dee\aoi
\;\geq\;
e^{c\epsilon}\int_{\epsilon}^{A}g(\aoi)\,\dee\aoi
\;\longrightarrow\;\infty
\qquad\text{as }c\to\infty \,.
\end{equation}
Since $\Lap{g}(0)=1<1/(\Rn\xp)$,
\cref{eq:bounded.support.Laplace} and continuity of $\Lap{g}$ show that
there is some $c>0$ for which
$\Lap{g}(-c)=1/(\Rn\xp)$.  Thus $r^{+}=-c$ is a negative root of
\cref{eq:rLaplace}, and the strict decrease of $\Lap{g}$ makes that
root unique.  Standard renewal-equation asymptotics then
identify $r^{+}$ with the falling-tail exponent $\lamp$
\citeref{Fell41}; see also
\citeref[Chapter~20, pp.\,353--364]{Kot2001}.  Thus $\lamp$ always exists
if the generation-interval density vanishes beyond some finite infection
age.

Bounded support of $g$ is sufficient but not necessary.  More generally, it
is enough that there is some $r_{*}<0$ for which $\Lap{g}(r_{*})$ is
finite and $\Lap{g}(r_{*})>1/(\Rn\xp)$.  Continuity of $\Lap{g}$ then
gives a root $r^{+}\in(r_{*},0)$ of \cref{eq:rLaplace}, and its strict
decrease makes that root unique.  Consequently,
\begin{equation}\label{eq:r.ineq}
  r_{*} < r^{+} < 0 < r^{-} \,.
\end{equation}
A density with unbounded support need not fail this condition;
the Gamma family in \cref{tab:GI.properties}, for instance, has
negative characteristic roots.  As a mathematical qualification,
however, a probability density with unbounded support can fail to
have any finite positive exponential moment, so that
$\Lap{g}(r)=\infty$ for every $r<0$.  Of course, a distribution with
unbounded support necessarily assigns positive probability to
transmission beyond every finite infection age, a condition that
cannot hold in a real biological system.

A mean-one lognormal generation-interval distribution with
log-variance $\nu>0$ provides a concrete example.  Let
\begin{equation}\label{eq:lognormal.g}
g(\aoi)
\;=\;
\frac{1}{\aoi\sqrt{2\pi\nu}}
\exp\!\left[-\frac{(\log\aoi+\nu/2)^{2}}{2\nu}\right],
\qquad \aoi>0 \,.
\end{equation}
This distribution has finite moments of every nonnegative
integer order.  For every $c>0$, however,
\begin{equation}\label{eq:lognormal.mgf}
\Lap{g}(-c)
\;=\;
\int_{0}^{\infty} e^{c\aoi}g(\aoi)\,\dee\aoi
\;=\;
\infty \,.
\end{equation}
Indeed, as $\aoi\to\infty$, the term $c\aoi$ in the logarithm
of the integrand dominates the terms of order $(\log\aoi)^{2}$.  Thus
$\lamm=r^{-}$ exists for this kernel, but no falling-tail exponent
$\lamp=r^{+}$ exists.  For the corresponding epidemic trajectory,
$Y(\tau)$, $\FoI(\tau)$ and $\inc(\tau)$ all decay more slowly than
$e^{-c\tau}$ for every $c>0$.

\beginsubappendix{Tail exponents for specific models}{app:tail.exponents}
\appfilename{epimom_theory/epimom_app_lampm_re.tex}

Given $g(\aoi)$, \cref{eq:rLaplace} or \cref{eq:lamLaplace} in
the main text characterizes $\lamm=r^{-}$ and $\lamp=r^{+}$.  When
either exponent has no explicit analytical
expression, it can be found by solving a one-dimensional implicit
equation.
\cref{tab:GI.properties} lists three simple cases for which
the expressions are explicit and simple, namely the SIR and SEIR ODE
models, and the renewal equation with a Gamma-distributed
$g(\aoi)$, which is commonly used because it looks similar to
observed generation interval distributions \citeref{DushPark21,Ferr+20}.
More realistic ODE models tend to yield cumbersome expressions for
$g(\aoi)$ if they are known (see, \eg Ref.~\citeref{Cham+18} for SEIR
models with Erlang-distributed latent and infectious periods).
 % app:lampm.re
\beginappendix{Initial epidemic momentum $\yinit$ from initial incidence $\incinit$}{app:yinit.from.incinit}
\appfilename{epimom_theory/epimom_app_yinit_from_incinit.tex}
\appbackrefs{\appbackref{finite-int-reps-main}{finite-incidence-history discussion}}

Under most circumstances, the first observed incidence occurs during the initial exponential growth phase \labelcref{eq:inc;tail}, whence for $\tau \leq \tauinit$,
\begin{equation}\label{eq:inc.init}
	\inc(\tau) \sim \incinit e^{\lamm (\tau-\tauinit)},
\end{equation}
In \appref[yinit-tail-exponents]{app:lampm.re} we observe that $\inc(\tau)$ and $Y(\tau)$ have the same initial exponential growth rate and, during the initial exponential phase, are approximately proportional, \labelcref{eq:Yinc}: $Y(\tau) \sim \yinit e^{\lamm (\tau - \tauinit)}$.  In particular, inserting the latter and \cref{eq:inc.init} into \cref{eq:Yinc}, we find
\begin{equation}\label{Yinc.init}
	\incinit \;\sim\; \frac{\yinit}{\int_{0}^{\infty} e^{-\lamm \aoi}
	\gisurv(\aoi)\, \dee \aoi}
	\;=\; \frac{\lamm \yinit}{1-\Lapm},
\end{equation}
where \cref{eq:Yinc.const} yields the latter equality.   Initial conditions for incidence, $\incinit$, and momentum, $\yinit$, are thus interchangeable.
 % app:yinit.from.incinit
\beginappendix{Existence of the momentum peak}{app:momentum.peak}
\appfilename{epimom_theory/epimom_app_momentum_peak.tex}
\appbackrefs{\appbackref{momentum-peak-existence}{momentum-peak statement}}

Recall that $\gisurv$ is the survival function of the intrinsic
generation interval distribution [\cref{eq:GI.survival}].
Because $g$ is a probability density, $0\leq\gisurv(\aoi)\leq1$,
$\gisurv$ is non-increasing, and
$\gisurv(\aoi)\to0$ as $\aoi\to\infty$.  The non-negative
initial data in \cref{eq:re.initial.FoI} imply that
$\FoI(\tau)\geq0$ for $\tau\geq\tauinit$, and \cref{eq:re;X}
therefore implies that $X$ remains non-negative and non-increasing.

Suppose that $X(\tauinit)>\xpeak$ and $\yinit>0$.  We first show that
epidemic momentum vanishes asymptotically.  For $\tau\geq\tauinit$,
splitting \cref{eq:re.Y.history} at $\tauinit$ gives
\begin{subequations}\label{eq:Y.asymptotic.split}
\begin{align}
  Y(\tau)
  &\;=\;
  Y_{\mathrm{hist}}(\tau)+Y_{\mathrm{new}}(\tau),
  \label{eq:Y.asymptotic.split;sum}
  \\
  Y_{\mathrm{hist}}(\tau)
  &\;=\;
  \int_0^\infty
    \inc_\init(\aoiinit)
    \gisurv(\tau-\tauinit+\aoiinit)\,\dee\aoiinit,
  \label{eq:Y.asymptotic.split;hist}
  \\
  Y_{\mathrm{new}}(\tau)
  &\;=\;
  \int_{\tauinit}^{\tau}
    \inc(s)\gisurv(\tau-s)\,\dee s .
  \label{eq:Y.asymptotic.split;new}
\end{align}
\end{subequations}
For every fixed $\aoiinit$, the integrand in
\cref{eq:Y.asymptotic.split;hist} tends to zero as $\tau\to\infty$,
and
\begin{equation}
  0
  \;\leq\;
  \inc_\init(\aoiinit)
  \gisurv(\tau-\tauinit+\aoiinit)
  \;\leq\;
  \inc_\init(\aoiinit)\gisurv(\aoiinit).
  \label{eq:Y.asymptotic.hist.bound}
\end{equation}
By \cref{eq:GI.survival,eq:re;yinit}, the bounding function
$\inc_\init(\aoiinit)\gisurv(\aoiinit)$ is integrable over
$[0,\infty)$, with integral $\yinit$.  Lebesgue's dominated
convergence theorem \citeref[\S11.32]{Rudin1976} therefore gives
\begin{equation}
  \lim_{\tau\to\infty}Y_{\mathrm{hist}}(\tau) \;=\; 0.
  \label{eq:Y.asymptotic.hist.limit}
\end{equation}

The incidence after $\tauinit$ also has finite integral.  Indeed,
$X(\tau)\to\xp$ as $\tau\to\infty$ [\cref{eq:xpm}], and
\cref{eq:re;X,eq:re;inc} give
\begin{equation}
  \int_{\tauinit}^{\infty}\inc(\taudum)\,\dee\taudum
  \;=\;
  \xinit-\xp
  \;\leq\;
  \xinit.
  \label{eq:post.initial.total.incidence}
\end{equation}
Given $\varepsilon>0$, choose $T\geq\tauinit$ so that
$\int_T^\infty\inc(s)\,\dee s<\varepsilon$.  Since $\gisurv$ is
non-increasing and bounded above by one, for $\tau>T$ we have
\begin{align}
  0
  \;\leq\;
  Y_{\mathrm{new}}(\tau)
  &\;\leq\;
  \gisurv(\tau-T)
  \int_{\tauinit}^{T}\inc(s)\,\dee s
  +
  \int_T^{\tau}\inc(s)\,\dee s
  \notag\\
  &\;<\;
  \gisurv(\tau-T)
  \int_{\tauinit}^{T}\inc(s)\,\dee s
  +\varepsilon .
  \label{eq:Y.asymptotic.new.bound}
\end{align}
Taking $\tau\to\infty$ and then $\varepsilon\to0$ shows that
$Y_{\mathrm{new}}(\tau)\to0$.  Together with
\cref{eq:Y.asymptotic.hist.limit}, this yields
\begin{equation}
  \lim_{\tau\to\infty}Y(\tau) \;=\; 0.
  \label{eq:Y.asymptotic.limit}
\end{equation}

If the trajectory never reached $X=\xpeak$, then $X(\tau)>\xpeak$ for
every $\tau\geq\tauinit$.  Since $\FoI(\tau)\geq0$,
\cref{eq:re.Ydot}
would then imply that $Y$ is non-decreasing and hence that
$Y(\tau)\geq\yinit>0$, contradicting
\cref{eq:Y.asymptotic.limit}.  The continuity of $X$ therefore implies
that there is a finite time $\taupeak$ at which
$X(\taupeak)=\xpeak$.  Since $X$ is non-increasing,
\cref{eq:re.Ydot} shows that $Y$ is non-decreasing before this crossing
and non-increasing afterwards.  Thus $Y$ attains its maximum for
$\tau\geq\tauinit$ when $X=\xpeak$.

\begin{samepage}
Under \cref{eq:re.initial.FoI}, \ie the \RE without external sources
of infection, all incidence after $\tauinit$ ultimately derives from
the initial history \labelcref{eq:re.initial.history}.  Thus $\yinit>0$
is also necessary
for $X(\tau)$ to cross $\xpeak$ in the future.  To see this, let
$B(\tau)$ denote the first integral in \cref{eq:re.initial.FoI}:
\begin{equation}
  B(\tau)
  \;=\;
  \int_0^\infty
    \inc_\init(\aoiinit)
    g(\tau-\tauinit+\aoiinit)\,\dee\aoiinit,
  \qquad \tau\geq\tauinit.
  \label{eq:history.future.FoI}
\end{equation}
Thus $\Rn B(\tau)$ is the contribution of the initial history to
$\FoI(\tau)$ in \cref{eq:re.initial.FoI}.
\end{samepage}
\begin{samepage}
Because $B$ is non-negative, it is positive on a set of positive
measure if and only if its integral
$\int_{\tauinit}^\infty B(\tau)\,\dee\tau$ is positive.  Substituting
\cref{eq:history.future.FoI} into this integral gives an iterated
integral with a non-negative integrand.  Tonelli's theorem
\citeref[Theorem~8.8(a), p.\,164]{Rudin1987} therefore permits us to
reverse the order of integration:
\begin{subequations}\label{eq:history.future.FoI.total}
\begin{align}
  \int_{\tauinit}^\infty B(\taudum)\,\dee\taudum
  &\;=\;
  \int_{\tauinit}^\infty\!\int_0^\infty
    \inc_\init(\aoiinit)
    g(\taudum-\tauinit+\aoiinit)
    \,\dee\aoiinit\,\dee\taudum
  \label{eq:history.future.FoI.total;original}
  \\
  &\;=\;
  \int_0^\infty
    \inc_\init(\aoiinit)
    \left[
      \int_{\tauinit}^\infty
      g(\taudum-\tauinit+\aoiinit)\,\dee\taudum
    \right]
    \dee\aoiinit
  \label{eq:history.future.FoI.total;reordered}
  \\
  &\;=\;
  \int_0^\infty
    \inc_\init(\aoiinit)\gisurv(\aoiinit)\,\dee\aoiinit
  \;=\;
  \yinit.
  \label{eq:history.future.FoI.total;value}
\end{align}
\end{subequations}
\end{samepage}\par\noindent
Changing variable in the inner integral in
\cref{eq:history.future.FoI.total;reordered} and using
\cref{eq:GI.survival} gives $\gisurv(\aoiinit)$ in
\cref{eq:history.future.FoI.total;value}; the final equality in
\cref{eq:history.future.FoI.total;value} is \cref{eq:re;yinit}.  Thus
$B$ is positive on a set of positive measure if and only if
$\yinit>0$.  In particular, if $\yinit=0$, then $B(\tau)=0$ for almost
every $\tau\geq\tauinit$, so the first integral in
\cref{eq:re.initial.FoI} vanishes.  That equation then reduces to a
homogeneous Volterra equation.  Its unique solution is
$\FoI(\tau)=0$ for $\tau\geq\tauinit$, so incidence is zero, $X$
remains at $\xinit$, and the trajectory cannot cross $\xpeak$.
Consequently, when
$\xinit>\xpeak$, positive initial epidemic momentum is both necessary
and sufficient for the trajectory to reach $X=\xpeak$.
 % app:momentum.peak
\beginappendix
    {Time transformation to map a general epidemic onto the SIR model}{app:timechange}
\appfilename{epimom_theory/epimom_app_timechange.tex}
\appbackrefs{\appbackref{timechange-sir-equivalence}{SIR time-transformation argument}}

Given \cref{eq:timechange}, the inverse function theorem and
the fundamental theorem of calculus imply that
\begin{align}
\ddtau{\Timetrans^{-1}}(\tau)
\;=\; \frac{1}{\Timetrans'(\Timetrans^{-1}(\tau))} \;=\; \frac{\Rn Y({\Timetrans^{-1}}(\tau)
)}{\FoI(\Timetrans^{-1}(\tau))}.
\end{align}
If we now define
%%\begin{subequations}
\begin{align}
\mathcal{X}(\tau) \;=\; X(\Timetrans^{-1}(\tau)) 
%%\noalign{\vspace{5pt}}
\,,\qquad
\mathcal{Y}(\tau) \;=\; Y(\Timetrans^{-1}(\tau)) ,
\end{align}
%%\end{subequations}
then the chain rule implies that
\begin{subequations}
\begin{align}
  \ddtau{\mathcal{X}} &\;=\; \ddtau{X(\Timetrans^{-1}(\tau))}
  \;=\; \dd{X}{\Timetrans^{-1}}\dd{\Timetrans^{-1}}{\tau}\\
  &\;=\; -X(\Timetrans^{-1}(\tau))\FoI(\Timetrans^{-1}(\tau)) \cdot \frac{\Rn Y({\Timetrans^{-1}}(\tau))}{\FoI(\Timetrans^{-1}(\tau))}  \;=\; -\Rn\mathcal{X}\mathcal{Y} \,, \nonumber\\
\intertext{and, similarly,}
\ddtau{\mathcal{Y}} &\;=\; (\Rn\mathcal{X}-1)\mathcal{Y} \,,
\end{align}
\end{subequations}
so $\mathcal{X}$ and $\mathcal{Y}$ satisfy the SIR equations \labelcref{eq:SIR}.
 % app:timechange

\beginappendix{\texorpdfstring{Lambert's $W$-function}{Lambert's W-function}}{app:LambertW}

\ifdefined\inferenceLambertWNote
  \inferenceLambertWNote
\fi

\appfilename{shared/epimom_app_lambertw.tex}
\appbackrefs{\appbackref{lambert-w}{Lambert $W$}}

If $\Winv(z) = ze^{z}$, Lambert's
$W$-function $W(z)$ 
(\citeref{Corl+96};
\citeref[\S4.13]{NIST:DLMF})
is the multivalued inverse of $\Winv$.  Its branches are denoted
$W_k(z)$, for integer $k$, and satisfy $\Winv(W_k(z))=z$ on their
respective domains.  We use the two real
branches: $W_{-1}$ maps $[-\frac{1}{e},0)$ to $(-\infty,-1]$, and
$W_{0}$ maps $[-\frac{1}{e},\infty)$ to $[-1,\infty)$.
Conversely, each branch $W_k$ is the inverse of $\Winv$ restricted to
the range of that branch.  In particular, on the real line, the
corresponding partial inverse relations are
\begin{subequations}\label{eq:Wid}
	\begin{align}
		W_{-1}(\Winv(z)) &\;=\; z \quad \text{if $z \leq -1$}\\
		W_{0}(\Winv(z)) &\;=\; z \quad \text{if $z \geq -1$}.
	\end{align}
\end{subequations}
While the standard notation $W_k$ is chosen to indicate
the winding number associated with the given branch, for our purposes
it is more convenient to write $\Wm$ for $W_{-1}$ and $\Wp$ for $W_0$,
so we can write expressions involving $\Wpm$, where the $\pm$ matches
the corresponding sign in $\xpm$ and/or $\lampm$ ($\Wp$ and $\Wm$ are 
also written ${\rm Wp}$ and ${\rm Wm}$ \citeref[\S4.13]{NIST:DLMF}).
 % app:LambertW
\beginappendix{\texorpdfstring{$\Rn\xm$ from $\lamp$}{R0 x- from lambda+}}{app:Rn.from.lamp}
\appfilename{epimom_theory/epimom_app_rn_from_lamp.tex}
\appbackrefs{\appbackref{rn-from-lamp-main}{falling-tail product relation}}

{\emergencystretch=2em
The endpoint products $\Rn\xm$ and $\Rn\xp$ are not independent:
because $(\xm,0)$ and $(\xp,0)$ lie on the same
susceptible--momentum trajectory, the first integral
\labelcref{eq:FI} gives
$\Volterra(\Rn\xm)=\Volterra(\Rn\xp)$.  Since
$\Rn\xm>1>\Rn\xp$ and $\Volterra$ is strictly monotone on either
side of 1, either endpoint product uniquely determines the other.
Taking the limit as the initial time approaches $-\infty$ in
\cref{eq:Xofy} (so $\xinit\to\xm$ and $\yinit\to0$) yields\par}
\begin{equation}\label{eq:Rnxp.of.Rnxm}
  \Rn\xpm \;=\; -\Wpm\big(\Winv(-\Rn\xmp)\big) .
\end{equation}
The falling-tail form of \cref{eq:lamLaplace;lam} gives
$\Rn\xp=1/\Lapp$.  Substituting this value into the minus case of
\cref{eq:Rnxp.of.Rnxm} yields
\begin{equation}\label{eq:Rn.via.Lapp}
\Rn\xm \;=\; -\Wm\big(\Winv(-\tfrac{1}{\Lapp})\big).
\end{equation}
{\emergencystretch=3em
Equation \labelcref{eq:Rn.via.Lapp} is the falling-tail
counterpart of $\Rn\xm=1/\Lapm$, the Euler--Lotka relationship
underlying the Wallinga--Lipsitch formula
\citeref[Equation~(2.7)]{WallLips07}.  Together, the rising-tail
relation $\Rn\xm=1/\Lapm$ and the falling-tail relation
\labelcref{eq:Rn.via.Lapp} connect the two tail exponents through their
common value $\Rn\xm$.\par}
 % app:Rn.from.lamp
\beginappendix{Extrema of prior immunity and final size as a function of \texorpdfstring{$\Rn$}{R0}}{app:zp.min}
\appfilename{epimom_theory/epimom_app_zpmin.tex}
\appbackrefs{\appbackref{zp-min-strict-inequalities}{strict-inequalities discussion};
\appbackref{zp-min-main}{nonmonotone final-size discussion}}

Evaluating \cref{eq:Yofx} at $\xpm$, we obtain the root equation
$Y(\xpm)=0$, \ie
\begin{equation}
  \yinit+\xinit-\xpm
  +\xpeak\log\!\Big(\frac{\xpm}{\xinit}\Big)
  \;=\;
  0.
  \label{eq:Yofx.root.equation}
\end{equation}
Implicit differentiation of \cref{eq:Yofx.root.equation} with respect
to $\xpeak$ gives
\begin{equation}
  \frac{\dee \xpm}{\dee \xpeak}
  \;=\;
  \frac{\xpm\log(\xpm/\xinit)}{\xpm-\xpeak}.
  \label{eq:dxpm.dxpeak}
\end{equation}
Since $\xpeak=\frac{1}{\Rn}$, we have
$\dd{\xpeak}{\Rn}=-\frac{1}{\Rn^2}=-\xpeak^2$, and therefore
\begin{equation}
  \frac{\dee}{\dee\Rn}(1-\xpm)
  \;=\;
  \xpeak^2
  \frac{\xpm\log(\xpm/\xinit)}{\xpm-\xpeak}.
  \label{eq:zpmin.endpoint.immunity.derivative}
\end{equation}
For $\xpm=\xm$, both $\log(\xm/\xinit)$ and $\xm-\xpeak$ are positive,
whereas for $\xpm=\xp$, both $\log(\xp/\xinit)$ and $\xp-\xpeak$ are
negative [\cf \cref{eq:xp.lt.xm}].  Hence the right-hand side of
\cref{eq:zpmin.endpoint.immunity.derivative} is positive for both
roots.  Increasing $\Rn$ therefore strictly increases both the
corresponding prior immunity, $\zm=1-\xm$,
and the total post-epidemic immune
fraction, $\zptot=1-\xp$.

The final size can be written as the difference of these two increasing
quantities:
\begin{equation}
  \zp
  \;=\;
  \xm-\xp
  \;=\;
  (1-\xp)-(1-\xm)
  \;=\;
  \zptot-\zm.
  \label{eq:zpmin.zp.difference.immunity}
\end{equation}
Consequently,
\begin{equation}
  \frac{\dee\zp}{\dee\Rn}
  \;=\;
  \frac{\dee\zptot}{\dee\Rn}
  -
  \frac{\dee\zm}{\dee\Rn}.
  \label{eq:zpmin.zp.derivative.difference}
\end{equation}
Thus $\zp$ decreases with $\Rn$ precisely when the
corresponding prior immunity $\zm$ increases
faster than the total post-epidemic immune
fraction $\zptot$.  It increases with $\Rn$ when the opposite
inequality holds.  This is why the fraction infected \emph{during} the
epidemic, $\zp(\Rn)$, can be non-monotone even though both endpoint
immune fractions are strictly increasing.

To locate critical points of $\zp(\Rn)$, we use
\cref{eq:Yofx.root.equation} to eliminate the logarithms in
\cref{eq:zpmin.endpoint.immunity.derivative} and then write
\begin{equation}\label{eq:zpmin.dzp.factorization}
  \frac{\dee\zp}{\dee\Rn} \;=\; \Phi_1(\Rn)\,\Phi_2(\Rn),
\end{equation}
where the factors $\Phi_1$ and $\Phi_2$ are functions of
$\xp,\xm,\xpeak$ (which are each functions of $\Rn$):
\begin{equation}\label{eq:zpmin.Phi1.Phi2}
  \Phi_1(\Rn)
  \;=\;
  \frac{\xpeak(\xm-\xp)}{(\xm-\xpeak)(\xpeak-\xp)}
  \,, \qquad
  \Phi_2(\Rn)
  \;=\;
  \xp\xm
  -
  \xpeak\big(\xp+\xm-\xinit-\yinit\big).
\end{equation}
Since $\xp<\xpeak<\xm$, the first factor $\Phi_1(\Rn)$ is strictly
positive.  Hence the critical points of $\zp(\Rn)$ are exactly the
zeros of $\Phi_2(\Rn)$, and a given critical point $\Rnstar$ is a
local minimum if and only if $\Phi_2$ is increasing at $\Rnstar$.

To express the local minimum condition ($\Phi_2(\Rn)$ increasing at
$\Rnstar$) in terms of $\Rnstar$, we first differentiate $\Phi_2$ with
respect to $\xpeak$:
\begin{equation}\label{eq:zpmin.Phi2.derivative.xpeak}
  \frac{\dee\Phi_2}{\dee\xpeak}
  \;=\;
  \frac{\dee\xp}{\dee\xpeak}\xm
  +
  \xp\frac{\dee\xm}{\dee\xpeak}
  -
  \big(\xp+\xm-\xinit-\yinit\big)
  -
  \xpeak\left(
    \frac{\dee\xp}{\dee\xpeak}
    +
    \frac{\dee\xm}{\dee\xpeak}
  \right).
\end{equation}
Then, using $\dd{\xpeak}{\Rn}=-\xpeak^2$, $\Phi_2(\Rnstar)=0$,
and \cref{eq:dxpm.dxpeak}, we
simplify \cref{eq:zpmin.Phi2.derivative.xpeak} at $\Rn=\Rnstar$ and
find
\begin{equation}
  \left.
  \frac{\dee\Phi_2}{\dee\Rn}
  \right|_{\Rn=\Rnstar}
  \;=\;
  -\xp\xm
  \big(\xp+\xm-\frac{3}{\Rnstar}\big).
  \label{eq:zpmin.Phi2.derivative.Rn.stationary}
\end{equation}
Therefore a critical point $\Rnstar$ is a local minimum of $\zp(\Rn)$
if and only if
\begin{equation}
  \xp(\Rnstar) \;+\; \xm(\Rnstar)
  \,\;<\;\,
  \frac{3}{\Rnstar} \,.
  \label{eq:zpmin.minimum.condition}
\end{equation}

If $\yinit=0$ and $\Rn=\frac{1}{\xinit}$ (or equivalently
$\xpeak=\xinit$) then $Y(x)$ has a double root ($\xp=\xm$) and hence
the final size $\zp=0$.  Thus $\Rn=\frac{1}{\xinit}$ is a global
minimum point of $\zp(\Rn)$ if $\yinit=0$.  By the implicit function
theorem, for sufficiently small $\yinit>0$, a local minimum in
$\zp(\Rn)$ near $\Rn=\frac{1}{\xinit}$ can be identified by perturbing
about $\Rn=\frac{1}{\xinit}$, or equivalently about $\xpeak=\xinit$.
If we set $x=\xinit+\delta$ in the root
equation \labelcref{eq:Yofx.root.equation} then we obtain
\begin{equation}\label{eq:zpmin.log.expansion.near.xinit}
  \yinit
  \;=\;
  \delta
  -\xinit\log\!\Big(1+\frac{\delta}{\xinit}\Big)
  \;=\;
  \frac{\delta^2}{2\xinit}
  +\Oh(\delta^3).
\end{equation}
Thus the two roots obtained with $\xpeak=\xinit$ can be written
$\xp=\xinit+\delta_-$ and $\xm=\xinit+\delta_+$, where
\begin{equation}
  \delta_\pm
  \;=\;
  \pm\sqrt{2\xinit\yinit}
  +\Oh(\yinit).
  \label{eq:zpmin.roots.near.xinit}
\end{equation}
Therefore, for $\Rn=1/\xinit$, the corresponding final size is
\begin{equation}
  \zp
  \;=\;
  \xm-\xp
  \;=\;
  \delta_+-\delta_-
  \;=\;
  \Oh(\sqrt{\yinit}).
  \label{eq:zpmin.small.upper.bound}
\end{equation}
Since the local minimum of $\zp(\Rn)$ in a neighbourhood of
$\Rn=1/\xinit$ cannot exceed the value obtained at $\Rn=1/\xinit$,
this local minimum final size tends to zero as $\yinit\to0$.

Expanding the two root equations for $\xp$ and $\xm$
[\cref{eq:Yofx.root.equation}], together with the critical point
condition ($\Phi_2(\Rnstar)=0$), in powers of $\yinit$ about the
double-root point,
\begin{equation}
  \yinit
  \;=\;
  0,
  \qquad
  \xpeakstar
  \;=\;
  \xinit,
  \qquad
  \xp
  \;=\;
  \xm
  \;=\;
  \xinit,
  \label{eq:zpmin.double.root.expansion.point}
\end{equation}
we find
\begin{equation}
  \frac{1}{\Rnstar}
  \;=\;
  \xpeakstar
  \;=\;
  \xinit-\yinit-\frac{7}{9\xinit}\yinit^2+\Oh(\yinit^3),
  \label{eq:zpmin.xpeakstar.expansion}
\end{equation}
which implies the leading approximation,
$\Rnstar\approx1/(\xinit-\yinit)$, as stated
in \cref{eq:zpmin.Rnstar.leading.approx}.

%%notes for derivation/approximation of maximum, which doesn't seem worth presenting:
%%\input{epimom_theory/epimom_app_zpmin_extra.tex}
 % app:zp.min
\beginappendix{Generalization of the removed compartment}{app:general.z}
\appfilename{epimom_theory/epimom_app_general_z.tex}
\appbackrefs{\appbackref{initial-conditions-charge}{charge and Z discussion}}

In analogy with KM's \citeref{KermMcKe27} SIR model, we can add a
third equation to \cref{eq:XYF},
\begin{equation}\label{eq:XYF;Z}
	\ddtau{Z} \;=\; \xpeak\FoI\,, \qquad\qquad Z(\tauinit) \;=\; \zinit.
\end{equation}
While not essential to describe the dynamics, \cref{eq:XYF;Z} is
potentially helpful conceptually.  Since the flow into $Z$ is equal to
the flow out of $Y$, we can think of $Z$ as an ``epidemic momentum
sink''.

The sum of \cref{eq:XYF;X,eq:XYF;Y,eq:XYF;Z} is $\ddtau{}(X+Y+Z) = 0$,
so $X+Y+Z$ is constant. Consequently, $\ddtau{Y} + \ddtau{Z} =
-\ddtau{X} = \inc$, the incidence, so
\begin{equation}\label{eq:YZ.incidence.history}
  Y(\tau) + Z(\tau)
  \;=\;
  \zm+\cuminc(\tau)
  \;=\;
  \zm +
  \int_0^\infty
    \inc(\tau-\aoi)\,\dee\aoi ,
\end{equation}
where $\zm=1-\xm$ is the pre-epidemic immune fraction and
$\cuminc(\tau)$ is the cumulative incidence during the epidemic up to
time $\tau$.  Since any individual who is not susceptible must either
be infected or removed, it follows that the constant value of $X+Y+Z$
is 1; in particular, $\xinit+\yinit+\zinit = 1$ so $\zinit$
in \cref{eq:XYF;Z} is determined by $(\xinit,\yinit)$.

We can also express $Z$ in terms of the incidence history.  Inserting
$Y$ from \cref{eq:re.Y} in \cref{eq:YZ.incidence.history} and
solving for $Z(\tau)$ yields
\begin{equation}
  Z(\tau)
  \;=\;
  \zm
  +
  \int_0^\infty
    \inc(\tau-\aoi)
    \Big(1-\gisurv(\aoi)\Big)
    \dee\aoi .
  \label{eq:re;Z}
\end{equation}
Thus $Y$ weights past incidence by its remaining reproductive potential,
whereas $Z$ weights past incidence by the reproductive potential already
consumed and adds this contribution to the pre-epidemic immune fraction
$\zm=1-\xm$.  Since both $\zm$ and the integral in \cref{eq:re;Z} are
non-negative, it follows that $Z(\tau)\ge0$ for all $\tau$.
 % app:general.z
\beginappendix{Time-varying reproduction number}{app:epimom.Rt}
\appfilename{epimom_theory/epimom_app_epimom_rt.tex}
\appbackrefs{\appbackref{epimom-rt-time-dependence}{time-dependent transmission discussion}}

Suppose that at time $\tau$, individuals of infectious age $\aoi$ give rise to new infections at rate $\beta(\tau,\aoi)$. Then the \term{instantaneous reproduction number, $\Rn(\tau)$} is \citeref{Fraser2007}
\begin{equation}\label{eq:Rt.def}
	\Rn(\tau) \;=\; \int_{0}^{\infty} \beta(\tau,\aoi)\, \dee \aoi.
\end{equation}
Following \citeref{Fraser2007},
we focus on the tractable case when $\beta(\tau,\aoi)$ is separable, \ie can
be written as the product of a function of $\tau$ and a function of $\aoi$,
which implies
\begin{equation}\label{eq:Rt.sep}
	 \beta(\tau,\aoi) \;=\; \Rn(\tau)g(\aoi),
\end{equation}
where, as in the main text, $g(\aoi)$ is the intrinsic generation
interval distribution.  In this situation,
$\Rn(\tau)$ can be factored out of an integral of $\beta$ with respect
to $\aoi$.

%%\beginsubappendix{Epidemic momentum with time-varying reproduction number}{subapp:Y.Rt}

Given \cref{eq:Rt.sep}, we can write the renewal equation \labelcref{eq:re} as
\begin{subequations}\label{eq:re.Rt}
\begin{align}
    \ddtau{X} &\;=\; - \inc(\tau) \,, \label{eq:re.Rt;X} \\
   \FoI(\tau) &\;=\; \Rn(\tau) \int_{0}^{\infty} \inc(\tau-\aoi)g(\aoi)\, \dee \aoi \,, \label{eq:re.Rt;FoI}
\end{align}
\end{subequations}
with $\inc(\tau)=X(\tau)\FoI(\tau)$, as usual.
Analogous to \cref{eq:Ra,eq:re.Y}, we can define
\begin{align}
  \Ra(\tau) &\;=\; \int_{\aoi}^{\infty} \Rn(\tau) \,g(\aoidum)\, \dee \aoidum
  \;=\;\Rn(\tau)\gisurv(\aoi) \,,
     \label{eq:Ra.Rt} \\
\text{and}\qquad
   Y(\tau) &\;=\; \int_{0}^{\infty} \inc(\tau-\aoi)
   \gisurv(\aoi)\, \dee \aoi
     \quad\;=\; \int_{0}^{\infty} \inc(\tau-\aoi) \int_{\aoi}^{\infty} \,g(\aoidum)\, \dee \aoidum\, \dee \aoi.
     \label{eq:re.Y.Rt}
\end{align}
Differentiating \cref{eq:re.Y.Rt}
under the integral sign we obtain
\begin{subequations}
\begin{align}
\ddtau{Y} &\;=\;  \int_{0}^{\infty} \ddtau{} \inc(\tau-\aoi)
  \gisurv(\aoi)\, \dee \aoi
   \;=\;  \int_{0}^{\infty} -\dd{}{\aoi} \inc(\tau-\aoi)
  \gisurv(\aoi)\, \dee \aoi
\intertext{so, integrating by parts,}
   &\;=\; -\inc(\tau-\aoi)
   \left.\gisurv(\aoi)
   \right|_{\aoi \;=\; 0}^{\aoi \;=\; \infty} - \int_{0}^{\infty} \!\inc(\tau-\aoi)\,g(\aoi)\, \dee \aoi\\
	&\;=\; \inc(\tau) - \frac{\FoI(\tau)}{\Rn(\tau)}
	 \quad\;=\; \Big(X(\tau)-\frac{1}{\Rn(\tau)}\Big)\FoI(\tau). \label{eq:dYdtau.Rt}
\end{align}
\end{subequations}
Thus the ODE for epidemic momentum $Y$ \labelcref{eq:XYF;Y} is
unchanged except that $\xpeak=\frac{1}{\Rn}$ is replaced by
$\frac{1}{\Rn(\tau)}$.
%%\beginsubappendix{Phase plane dynamics with time-varying reproduction number}{subapp:pp.Rt}
However,
while the epidemic momentum is well-defined with time-varying $\Rn$,
dividing \cref{eq:dYdtau.Rt} by \cref{eq:re.Rt;X} replaces \cref{eq:dYdx} with the non-autonomous equation
\begin{equation}\label{eq:dYdx.Rt}
\ddx{Y} \;=\; -1 + \frac{1}{\Rn(\tau)\,x} \,, \qquad Y(\xinit) \;=\; \yinit,
\end{equation}
which has no simple generic solution.
 % app:epimom.Rt

%% Appendices for which the citation path does not end in epimom_main.tex, so are not actually used anywhere:
%%%%\input{epimom_theory/epimom_app_inc_from_epimom} % app:inc.from.epimom
%%%%\input{epimom_theory/epimom_app_re_from_y} % app:RE.from.Y
%%%%\input{epimom_theory/epimom_app_final_size_rescaling} % app:final.size.rescaling

\bibliographystyle{RS}
\bibliography{epimom-refs}

\end{document}